\documentclass[11pt]{article}

\usepackage[margin=.75in]{geometry}

\usepackage{amsmath,amssymb}
\usepackage{graphicx}
\usepackage{caption}
\usepackage{subcaption}
\usepackage{color}
\definecolor{indigo}{RGB}{0,0,120}
\usepackage[colorlinks=true, linkcolor=blue, citecolor=red, urlcolor=indigo]{hyperref}
\usepackage{comment}

\def\tr{\,{\rm tr}\,}
\def\Tr{\,{\rm Tr}\,}

\def\fl{\noindent}

\newcommand{\tl}[1]{\tilde{#1}}
\newcommand{\dd}[2]{\frac {\partial #1}{\partial #2}}
\newcommand{\deldel}[2]{\frac {\delta #1}{\delta #2}}
\newcommand{\pdr}{\partial}
\newcommand{\DD}[2]{\frac {d #1}{d #2}}

\newcommand{\beq}{\begin{equation}}
\newcommand{\eeq}{\end{equation}}
\newcommand{\beqs}{\begin{eqnarray}}
\newcommand{\eeqs}{\end{eqnarray}}
\newcommand{\half}{\frac{1}{2}}
\newcommand{\ov}[1]{\frac{1}{#1}}
\newcommand{\sign}[1]{{\rm sgn}({#1})}

\def\al{\alpha} 		
\def\del{\delta}
\def\D{\Delta}	
\def\g{\gamma} 
\def\eps{\epsilon} 
\def\veps{\varepsilon} 
\def\la{\lambda}
\def\La{\Lambda}		
\def\sig{\sigma}
\def\Sig{\Sigma}
\def\tht{\theta}
		
\def\Om{\Omega}

\newcount\colveccount  
\newcommand*\colvec[1]{\global\colveccount#1  \begin{pmatrix} \colvecnext} \def\colvecnext#1{#1 \global\advance\colveccount-1
        \ifnum\colveccount>0 \\ \expandafter\colvecnext
        \else \end{pmatrix} \fi}

\usepackage{calligra} 
\DeclareMathAlphabet{\mathcalligra}{T1}{calligra}{m}{n}
\DeclareFontShape{T1}{calligra}{m}{n}{<->s*[2.2]callig15}{}
\newcommand{\scripty}[1]{\ensuremath{\mathcalligra{#1}}}

\begin{document}


\title{
\hfill {\tt \small \href{https://arxiv.org/abs/1804.02859}{arXiv:1804.02859 [hep-th]}} \\
On the Hamiltonian formulation, integrability and algebraic structures of the Rajeev-Ranken model}
\author{{\sc Govind S. Krishnaswami and T. R. Vishnu}
\\ \small
Chennai Mathematical Institute,  SIPCOT IT Park, Siruseri 603103, India
\\ \small
Email: {\tt govind@cmi.ac.in, vishnu@cmi.ac.in}}
\date{February 6, 2019 \\ 
Published in \href{https://doi.org/10.1088/2399-6528/ab02a9}{J. Phys. Commun. {\bf 3}, 025005 (2019)}
}
\maketitle

\abstract{ \small The integrable 1+1-dimensional SU(2) principal chiral model (PCM) serves as a toy-model for 3+1-dimensional Yang-Mills theory as it is asymptotically free and displays a mass gap. Interestingly, the PCM is `pseudodual' to a  scalar field theory introduced by Zakharov and Mikhailov and Nappi that is strongly coupled in the ultraviolet and could serve as a toy-model for non-perturbative properties of theories with a Landau pole. Unlike the `Euclidean' current algebra of the PCM, its pseudodual is based on a nilpotent current algebra. Recently, Rajeev and Ranken obtained a mechanical reduction by restricting the nilpotent scalar field theory to a class of constant energy-density classical waves expressible in terms of elliptic functions, whose quantization survives the passage to the strong-coupling limit. We study the Hamiltonian and Lagrangian formulations of this model and its classical integrability from an algebraic perspective, identifying Darboux coordinates, Lax pairs, classical $r$-matrices and a degenerate Poisson pencil. We identify Casimirs as well as a complete set of conserved quantities in involution and the canonical transformations they generate. They are related to Noether charges of the field theory and are shown to be generically independent, implying Liouville integrability. The singular submanifolds where this independence fails are identified and shown to be related to the static and circular submanifolds of the phase space. We also find an interesting relation between this model and the Neumann model allowing us to discover a new Hamiltonian formulation of the latter.}

\vspace{.2cm}


{\footnotesize \tableofcontents}

\normalsize

\section{Introduction}

It is well-known that the 1+1-dimensional SU(2) non-linear sigma model (NLSM) and the closely related principal chiral model (PCM) for the SU(2)-valued field $g(x,t)$ are good toy-models for the physics of the strong interactions and 3+1-dimensional Yang-Mills theory. They have been shown to be asymptotically free and to possess a mass-gap \cite{Polyakov}. Non-perturbative results concerning the $S$-matrix and the spectrum of the 1+1- dimensional NLSM and PCM have been obtained using the methods of integrable systems by Zamolodchikov and Zamolodchikov \cite{Z-Z} (factorized $S$-matrices), by Polyakov and Wiegmann \cite{P-W} (fermionization) and by Faddeev and Reshetikhin \cite{F-R} (quantum inverse scattering method). Interestingly, a `pseudodual' to the PCM introduced in the work of Zakharov and Mikhailov \cite{Z-M} and Nappi \cite{Nappi} is strongly coupled in the ultraviolet, displays particle production and has been shown by Curtright and Zachos \cite{C-Z} to possess infinitely many {\it non-local} conservation laws. Thus, this dual scalar field theory could serve as a toy-model for studying certain non-perturbative aspects of 3+1-dimensional $\la \phi^4$ theory which appears in the scalar sector of the standard model. 

Before proceeding with our discussion of this dual scalar field theory, it is interesting to note that variants of this model, their integrability and the pseudoduality transformation have been investigated in various other contexts. For instance, a generalization to a centrally-extended Poincar\'e group leads to a model for gravitational plane waves \cite{N-W}. On the other hand, a generalization to other compact Lie groups shows that the pseudodual models have 1-loop beta functions with opposite signs \cite{Alvarez}. Interestingly, the sigma model for the non-compact Heisenberg group is also closely connected to the above dual scalar field theory \cite{B-Y}. Similar duality transformations have also been employed in the $AdS_5 \times S^5$ superstring sigma model in connection with the Pohlmeyer reduction \cite{G-T} and in integrable $\la$-deformed sigma models \cite{G-S-S}. The above dual scalar field theory also arises in a large-level and weak-coupling limit of the Wess-Zumino-Witten model and is also of interest in connection with the theory of hypoelliptic operators \cite{R-R}. In another direction, attempts have been made to understand the connection (or lack thereof) between the absence of particle production, integrability and factorization of the tree-level S-matrix in massless 2-dimensional sigma models \cite{H-L-T}.

Returning to the SU(2) principal chiral model, we recall that it is based on the semi-direct product of an $\mathfrak{s u}(2)$ current algebra and an abelian algebra (`Euclidean' current algebra) \cite{F-T}. On the other hand, its dual is based on a step-3 nilpotent algebra of currents $I = g^{-1}g'/\la^2$ and $J = g^{-1} \dot{g}/\la$, where $\la$ is a dimensionless coupling constant (see Eq.~(\ref{e:PB-currents})). Systems admitting a formulation based on quadratic Hamiltonians and nilpotent Lie algebras are particularly interesting, they include the harmonic and anharmonic oscillators as well as field theories such as $\la \phi^4$, Maxwell and Yang-Mills \cite{R-R}. Interestingly, the equation of motion (EOM) of the PCM $(\dot J = \la I')$ can be solved by expressing the currents $I = \dot \phi/ \la$ and $J = \phi'$ in terms of an $\mathfrak{s u}(2)$-valued scalar field $\phi(x,t)$. The zero-curvature consistency condition $(\dot I - J'/\la = \la [I, J])$ then becomes a non-linear wave equation: 
	\beq
	\ddot \phi = \phi'' + \la [ \dot{\phi}, \phi' ].
	\label{e:Non-linear-wave-equations}
	\eeq 
Recently, Rajeev and Ranken \cite{R-R} studied a class of constant energy-density `continuous wave' solutions to (\ref{e:Non-linear-wave-equations}) obtained via the ansatz
	\beq
	\phi(x,t) = e^{Kx} R(t) e^{-Kx} + m K x \quad \text{where} \quad K = \frac{i k \sigma_3}{2}
	\eeq
and $R(t)$ is a traceless $2 \times 2$ anti-hermitian matrix. The continuous waves depend on two constants, a wavenumber $k$ and a dimensionless parameter $m$. The reduction of the nilpotent scalar field theory to the manifold of these continuous waves is a mechanical system, the `Rajeev-Ranken' (RR) model, with three degrees of freedom $R_a = \Tr(R \sigma_a/ 2i)$ where $\Tr X = -2 \tr X$. Interestingly, the continuous wave  solutions remain non-trivial even in the limit of strong coupling so that their quantization could play a role in understanding the microscopic degrees of freedom of the corresponding quantum theory. In \cite{R-R}, conserved quantities of the RR model were used to reduce the EOM for $R(t)$ to a single non-linear ODE which was solved in terms of the Weierstrass $\wp$ function. 

In this article, we study the classical dynamics of the RR model focussing on its Hamiltonian formulation and aspects of its integrability especially through its algebraic structures. We begin by reviewing the passage from the PCM to the nilpotent scalar field theory, followed by its reduction to the RR model in sections \ref{s:PCM-NPF-RR} and \ref{s:NFT-to-RR}. Just as the canonical Poisson brackets (PBs) between  $I$ and its conjugate momentum in the Lagrangian of the PCM lead to the Euclidean Poisson algebra among currents $I$ and $J$ \cite{F-T}, the canonical PBs between $\phi$ and its conjugate momentum are shown to imply a step-3 nilpotent Poisson algebra among these currents. In section \ref{s:Darboux-coordinates-M6}, we identify canonical Darboux coordinates $(R_a, kP_a)$ on the six-dimensional phase space of the RR model and a Hamiltonian formulation thereof. These coordinates are used to deduce a Lagrangian formulation, as a naive reduction of the field theoretic Lagrangian does not do the job. Interestingly, since the evolution of $R_3$ decouples from that of the remaining variables, it is possible to give an alternative Hamiltonian formulation in terms of the variables $L = \left[ K, R \right] + mK $ and $S = \dot{R} + K/\la $ introduced by Rajeev and Ranken (see section \ref{s:Hamiltonian-mechanical}). The latter include a non-dynamical constant $L_3 = -mk$ but have the advantage of satisfying a step-3 nilpotent Poisson algebra which may be regarded as a finite dimensional version of the current algebra of the scalar field theory. Remarkably, the EOM in terms of the $S$ and $L$ variables admit another Hamiltonian formulation with the same Hamiltonian but PBs that are a finite dimensional analogue of the Euclidean current algebra of the PCM. Moreover, the nilpotent and Euclidean Poisson structures are compatible and combine to form a Poisson pencil as shown in section \ref{s:semi-direct-product-PB}. However, all the resulting Poisson structures are degenerate so that this Poisson pencil does not lead to a bi-Hamiltonian structure. In section \ref{s:Lax-pair-and-r-matrix}, we find Lax pairs and classical $r$-matrices with respect to both Poisson structures and use them in section \ref{s:conserved-quantities} to identify a maximal set of four conserved quantities in involution ($\mathfrak{c}, m, s^2$ and $h$). These conserved quantities are quadratic polynomials in $S$ and $L$. While $\mathfrak{c}$ and $m$ are Casimirs of the nilpotent $S$-$L$ Poisson algebra, $s^2$ and $h$ are Casimirs of the Euclidean Poisson algebra. While $h k^2 = \Tr SL$ is loosely like helicity, the Hamiltonian is proportional to $s^2 k^2 = \Tr S^2$ upto the addition of a term involving $\mathfrak{c}$. In section \ref{s:Noether-symmetries-CT}, we find the canonical transformations generated by these conserved quantities and the associated symmetries. In section \ref{s:conserved-qtys-Noether-charges} we also relate three of the conserved quantities to the reduction of Noether charges of the field theory. In section \ref{s:independence}, we show that the conserved quantities are generically independent and (a) identify submanifolds of the phase space where this independence fails and (b) the corresponding relations among conserved quantities. We also discover that these singular submanifolds are precisely the places (found in \S \ref{s:Static-and-trigonometric-solutions}) where the equations of motion may be solved in terms of circular rather than elliptic functions. The independence and involutive property of the conserved quantities imply Liouville integrability of the RR model \cite{Arnold}. Interestingly, we also find a mapping of variables that allows us to relate the EOM and Lax pairs of the RR model to those of the Neumann model \cite{B-T,B-B-T}. In section \ref{s:RR-model-and-Neumann-Model} this map is used to propose a new Hamiltonian formulation of the Neumann model with a nilpotent Poisson algebra. Despite some similarities between the models, there are differences: while $P$ and $J$ in the Neumann model are a projection and a real anti-symmetric matrix, the corresponding $S$ and $L$ variables of the RR model are anti-hermitian, so that the Poisson structures as well as $r$ matrices of the two models are distinct. We conclude with a brief discussion in section \ref{s:Discussion}. 

\section{From the SU(2) PCM to the nilpotent scalar field theory}
\label{s:PCM-NPF-RR}

The 1+1-dimensional principal chiral model is defined by the action
	\beq
	S_{\rm PCM} = \frac{1}{2\la^2} \int \Tr \left(\pdr_{\mu}g \pdr^{\mu}g^{-1}\right) dx dt 
	= \frac{1}{2\la^2} \int \Tr \left[ (g^{-1}\dot g )^2 - (g^{-1}g')^2 \right] dx dt,
	\label{e:action-PCM}
	\eeq
with primes and dots denoting $x$ and $t$ derivatives. Here, $\la > 0$ is a dimensionless coupling constant and $\Tr = -2 \tr$. The corresponding equations of motion (EOM) are non-linear wave equations for the components of the SU(2)-valued field $g$ and may be written in terms of the Lie algebra-valued time and space components of the right current, $r_0 = g^{-1}\dot{g}$ and $r_1 = g^{-1}g'$:
	\beq
	\ddot g - g'' = \dot g g^{-1} \dot g - g' g^{-1}g' \qquad \text{or} \qquad	
	\dot r_0 - r_1' = 0.
	\eeq
An equivalent formulation is possible in terms of left currents $l_{\mu} = (\partial_{\mu} g) g^{-1}$. Note that $r_0$ and $r_1$ are components of a flat connection; they satisfy the zero curvature `consistency' condition
	\beq
	\dot{r_1} - r_0' + \left[r_0, r_1\right] = 0.
	\eeq
Following Rajeev and Ranken \cite{R-R}, we define right current components rescaled by $\la$, which are especially useful in discussions of the strong coupling limit:
	\beq
	I = \frac{1}{\la^2}r_1 \quad \text{and} \quad J = \frac{1}{\la}r_0.
	\eeq
In terms of these currents, the EOM and zero-curvature condition become 
	\beq
	\dot{J} = \la I' \quad \text{and} \quad
	\dot{I} = \la\left[I,J\right] +\frac{1}{\la} J'.
	\label{e:eom-ZC-currents}
	\eeq
These EOM may be derived from the Hamiltonian following from $S_{\rm PCM}$ (upon dividing by $\la$),
	\beq
	H_{\rm PCM}  = \half \Tr \int dx \left(\la I^2 + \ov{\la} J^2 \right) 
	\label{e:H-PCM} 
	\eeq
and the PBs:
	\beqs
	\{ I_a(x), I_b(y) \} &=& 0, \quad \{ J_a(x), J_b(y) \} = - \la^2 \eps_{abc} J_c(x) \del(x-y) \cr
	\qquad \text{and} \quad \{ J_a(x), I_b(y) \} &=& -\la^2 \eps_{abc} I_c(x) \del(x-y) + \del_{ab} \pdr_x \del(x-y) \quad \text{for} \quad a,b = 1,2,3.
	\label{e:PB-PCM} 
	\eeqs
Since both $I$ and $J$ are anti-hermitian, their squares are negative operators, but the minus sign in $\Tr$ ensures that $H_{\rm PCM} \geq 0$. The Poisson algebra (\ref{e:PB-PCM}) is a central extension of a semi-direct product of the abelian algebra generated by the $I_a$ and the $\mathfrak{su}(2)$ current algebra generated by the $J_a$. It may be regarded as a (centrally extended) `Euclidean' current algebra. These PBs follow from the canonical PBs between $I$ and its conjugate momentum in  the  action (\ref{e:action-PCM}) \cite{F-T}. The multiplicative constant in $\{ J_a, J_b \}$ is not fixed by the EOM. It has been chosen for convenience in identifying Casimirs of the reduced mechanical model in \S \ref{s:semi-direct-product-PB}. 

The EOM $\dot J = \la I'$ is identically satisfied if we express the currents in terms of a Lie algebra-valued potential $\phi$: 
	\beq
	I = \frac{\dot \phi}{\la} 
	\quad \text{and} \quad J = \phi' \quad \text{or} \quad 
	r_{\mu} = \la \eps_{\mu \nu} \partial^{\nu} \phi \quad \text{with} \quad 
	g_{\mu \nu} = \colvec{2}{1 & 0 }{0 & -1} \quad \text{and} \quad \eps^{0 1} = 1.	
	\label{e:currents-NFT}
	\eeq
The zero curvature condition ($\dot{I} -  J'/ \la = \la\left[I,J\right]$) now becomes a $2^{\rm nd}$-order non-linear wave equation for the scalar $\phi$ (with the speed of light re-instated):
	\beq
	\ddot\phi = c^2 \phi'' + c \la [ \dot{\phi}, \phi' ].
	\label{e:Non-linear-wave-equation}
	\eeq
The field $\phi$ is an anti-hermitian traceless $2 \times 2$ matrix in the $\mathfrak{su}(2)$ Lie algebra, which may be written as a linear combination of the generators $t_a = \sigma_a / 2i$ where $\sig_a$ are the Pauli matrices:
	\beq
	\phi = \phi_a t_a = \frac{1}{2i} \phi \cdot \sigma 
	\quad \text{with} \quad \phi_a = i \tr(\phi \sigma_a) = \Tr(\phi t_a) \quad \text{for} \quad a = 1, 2, 3.
	\eeq 
The generators are normalized according to $\Tr(t_a t_b) = \del_{ab}$ and satisfy $\left[ t_a , t_b \right] = \eps_{abc} t_c$. As noted in \cite{R-R}, a strong-coupling limit of (\ref{e:Non-linear-wave-equation}) where the $\la [\dot \phi, \phi']$ term dominates over $\phi''$, may be obtained by introducing the rescaled field $\tl \phi (\xi, \tau) = \la^{2/3} \phi(x,t)$, where $\xi = x$ and $\tau = \la^{1/3} t$. Taking $\la \to \infty$ holding $c$ fixed gives the Lorentz non-invariant equation $\tl \phi_{\tau \tau} = c[\tl \phi_{\tau} , \tl \phi_{\xi}]$. Contrary to the expectations in \cite{R-R}, the `slow-light' limit $c \to 0$ holding $\la$ fixed is not quite the same as this strong-coupling limit.

The wave equation (\ref{e:Non-linear-wave-equation}) follows from the Lagrangian density (with $c=1$)
	\beq
	\mathcal{L} = \Tr \left( \frac{1}{2\la}(\dot{\phi}^2 - \phi'^2) + \frac{1}{3}\phi [\dot{\phi}, \phi'] \right) 
	 = \ov{2\la} \pdr_{\mu} \phi_a\pdr^{\mu} \phi_a + \frac{1}{6} \eps_{abc}\epsilon^{\mu \nu}\phi_a \pdr_{\mu} \phi_b \pdr_{\nu}\phi_c.
	\label{e:Lagrangian-scalar-field}
	\eeq 

The momentum conjugate to $\phi$ is $\pi =  \dot \phi/ \la  - (1/3) \left[\phi , \phi'\right]$ and satisfies 
		\beq
	\dot \pi = \frac{\phi''}{\la} + \frac{2}{3} [ \dot \phi, \phi' ] + \frac{1}{3} [\dot \phi' , \phi ] 
	= \frac{\phi''}{\la} + \frac{2 \la }{3} [ \pi , \phi' ] + \frac{\la}{3}[ \pi' , \phi ] + \frac{2 \la}{9} [ [ \phi , \phi'], \phi'] + \frac{\la}{9} [[\phi , \phi''], \phi].
	\label{e:pi-field}
	\eeq
The conserved energy and Hamiltonian coincide with $H_{\rm PCM}$ of (\ref{e:H-PCM}):
	\beq
	E = \frac{1}{2\la} \Tr \int dx \left[ \dot \phi^2 + \phi'^2 \right] 
	\quad \text{and} \quad
	H  = \half \Tr \int  dx\: \left[ \la \left(\pi + \frac{1}{3}[ \phi, \phi']\right)^2 + \ov \la \phi'^2\right]. 
	\label{e:H-Nilpotent-field-theory}
	\eeq
If we postulate the canonical PBs
	\beq
	\{ \phi_a(x), \phi_b(y) \} = 0, \quad \{ \phi_a(x), \pi_b(y) \} = \del_{ba} \del(x-y) \quad \text{and} \quad \{ \pi_a(x) , \pi_b(y) \} = 0,
	\label{e: Canonical-PB-field}
	\eeq
then Hamilton's equations $\dot \phi = \{ \phi , H \}$ and $\dot \pi = \{ \pi , H \}$ reproduce (\ref{e:pi-field}).
The canonical PBs between $\phi$ and $\pi$ imply the following PBs among the currents $I, J$ and $\phi$:
	\beqs
	\{ J_a(x) , J_b(y) \} = 0, \quad \{ I_a(x), J_b(y) \} &=&  \delta_{a b} \pdr_x \delta(x-y), \quad 
	\{ \phi_a(x) , I_b(y) \} =  \del_{ab} \del (x-y), \cr
	\{ \phi_a(x), J_b(y) \} = 0 \quad\text{and} \quad \{ I_a(x), I_b(y) \} &=&  \frac{\eps_{abc}}{3} \left( 2 J_c(x) + (\phi_c(x) - \phi_c(y)) \pdr_y \right) \del(x-y).
	\label{e:PB-IJ-phi}
	\eeqs
These PBs define a step-3 nilpotent Lie algebra in the sense that all triple PBs such as 
	\beq
	\{ \{ \{I_a(x), I_b(y)\} , I_c(z)\}, I_d(w) \}
	\eeq
vanish. Note however that the currents $I$ and $J$ do not form a closed subalgebra of (\ref{e:PB-IJ-phi}). Interestingly, the EOM (\ref{e:eom-ZC-currents}) also follow from the same Hamiltonian (\ref{e:H-PCM}) if we postulate the following closed Lie algebra among the currents
	\beq
	\{ J_a(x) , J_b(y) \} = 0, \;\;
	\{ I_a(x), J_b(y) \} =  \delta_{a b} \pdr_x \delta(x-y) \;\; \text{and} \;\;
	\{ I_a(x) , I_b(y) \} = \epsilon_{abc}J_c\delta(x-y).
	\label{e:PB-currents}
	\eeq
Crudely, these PBs are related to (\ref{e:PB-IJ-phi}) by `integration by parts'. As with (\ref{e:PB-IJ-phi}), this Poisson algebra of currents is a nilpotent Lie algebra of step-3 unlike the Euclidean algebra of Eq.~(\ref{e:PB-PCM}).

The scalar field with EOM (\ref{e:Non-linear-wave-equation}) and Hamiltonian (\ref{e:H-Nilpotent-field-theory}) is classically related to the PCM through the change of variables $r_{\mu} = \la \eps_{\mu \nu} \partial^{\nu} \phi$. However, as noted in \cite{C-Z}, this transformation is not canonical, leading to the moniker `pseudodual'. Though this scalar field theory has not been shown to be integrable, it does possess infinitely many (non-local) conservation laws \cite{C-Z}. Moreover, the corresponding quantum theories are different. While the PCM is asymptotically free, integrable  and serves as a toy-model for 3+1D Yang-Mills theory, the quantized scalar field theory displays particle production (a non-zero amplitude for $2 \to 3$ particle scattering), has a positive $\beta$ function \cite{Nappi} and could serve as a toy-model for 3+1D $\la \phi^4$ theory \cite{R-R}.

\section{Reduction of the nilpotent field theory and the RR model}
\label{s:NFT-to-RR}

Before attempting a non-perturbative study of the nilpotent field theory, it is interesting to study its reduction to finite dimensional mechanical systems obtained by considering special classes of solutions to the non-linear wave equation (\ref{e:Non-linear-wave-equation}). The simplest such solutions are traveling waves $\phi(x,t) = f(x-vt)$ for constant $v$. However, for such $\phi$, the commutator term $ - \la [v f', f'] = 0$ so that traveling wave solutions of (\ref{e:Non-linear-wave-equation}) are the same as those of the linear wave equation. Non-linearities play no role in similarity solutions either. Indeed, if we consider the scaling ansatz $\tl \phi\left(\xi, \tau \right) = \La^{-\gamma} \phi(x,t)$ where $\xi = \La^{-\alpha} x$ and $\tau = \La^{-\beta}t$, then (\ref{e:Non-linear-wave-equation}) takes the form:
	\beq
	\La^{\gamma - 2\beta} \tl \phi_{\tau \tau} - \La^{\gamma - 2 \alpha} \tl \phi_{\xi \xi} - \La^{2 \gamma - (\beta + \alpha)} \la [ \tl \phi_{\tau} , \tl \phi_{\xi} ] = 0. 
	\eeq
This equation is scale invariant when $\alpha = \beta$ and $\gamma = 0$. Hence similarity solutions must be of the form $\phi(x,t) =  \psi(\eta)$ where $\eta = x/t$ and $\psi$ satisfies the {\it linear} ODE
	\beq
	\eta^2 \psi'' - \psi'' + 2 \eta \psi' = - \la \eta [ \psi', \psi'] = 0.	
	\eeq

Recently, Rajeev and Ranken \cite{R-R} found a mechanical reduction of the nilpotent scalar field theory for which the non-linearities play a crucial role. They considered the wave ansatz:
	\beq
	\phi(x,t) = e^{Kx}R(t)e^{-Kx} + mKx\quad \text{with} \quad 
	K = \frac{i}{2} k \sigma_3
	\label{e:ansatz}
	\eeq
which leads to `continuous wave' solutions of (\ref{e:Non-linear-wave-equation}) with constant energy density.  These screw-type configurations are obtained from a Lie algebra-valued matrix $R(t)$ by combining an internal rotation (by angle $\propto x$) and a translation. The constant traceless anti-hermitian matrix $K$ has been chosen in the $3^{\rm rd}$ direction. The ansatz (\ref{e:ansatz}) depends on two parameters: a dimensionless real constant $m$ and the constant $K_3 = -k$ with dimensions of a wave number which could have either sign. When restricted to the submanifold of such propagating waves, the field equations (\ref{e:Non-linear-wave-equation}) reduce to those of a mechanical system with 3 degrees of freedom which we refer to as the Rajeev-Ranken model. The currents (\ref{e:currents-NFT}) can be expressed in terms of $R$:
	\beq
	I = \frac{1}{\la}e^{Kx}\dot{R}e^{-Kx} \quad \text{and} \quad 
	J = e^{Kx} \left( {\left[K,R\right] +mK} \right) e^{-Kx}.
	\label{e:currents-I-J}
	\eeq
These currents are periodic in $x$ with period $2\pi/|k|$. We work in units where $c=1$ so that $I$ and $J$ have dimensions of a wave number. If we define the traceless anti-hermitian matrices
	\beq
	L = \left[ K, R \right] + mK
	\quad \text{and}
	\quad S = \dot{R} + \ov{\la} K,
	\label{e: L-and-S}
	\eeq
then it is possible to express the EOM and consistency condition (\ref{e:eom-ZC-currents}) as the pair
	\beq
	\dot{L} = \left[K, S\right] 
	\quad \text{and} \quad 
	\dot{S} = \la \left[S, L\right].
	\label{e: EOM-LS}
	\eeq
In components $(L_a = \Tr( L t_a)$ etc.), the equations become 
	\beqs
	\dot L_1 &=& k S_2, \qquad \dot L_2 = - k S_1, \qquad \dot L_3 = 0, \cr 
	\dot S_1 &=& \la (S_2 L_3 - S_3 L_2),
	\quad \dot S_2 = \la (S_3 L_1 - S_1 L_3) \quad \text{and} \quad
	\;\; \dot S_3 = \la (S_1 L_2 - S_2 L_1).
	\label{e:EOM-LS-explicit}
	\eeqs
Here, $L_3 = - m k$ is a constant, but it will be convenient to treat it as a coordinate. Its constancy will be encoded in the Poisson structure so that it is either a conserved quantity or a Casimir. Sometimes it is convenient to express $L_{1,2}$ and $S_{1,2}$ in terms of polar coordinates:
	\beq
	L_1 = kr \cos \tht, \quad L_2 = kr \sin \tht, \quad S_1 = k \rho \cos \phi \quad \text{and} \quad S_2 = k \rho \sin \phi.
	\label{e:L-S-polar}
	\eeq 
Here, $r$ and $\rho$ are dimensionless and positive. We may also express $L$ and $S$ in terms of coordinates and velocities (here $u = \dot{R_3}/k - 1/\la$):
	\beqs
	L &=& \frac{k}{2i} \colvec{2}{-m & R_2 + i R_1}{R_2 - iR_1 & m}
\quad \text{and} \quad 
	S = \frac{1}{2i} \colvec{2}{ uk & \dot{R_1} - i \dot{R_2}}{\dot{R_1} + i\dot{R_2} & -uk} \quad \text{or} \cr
	L_1 &=& k R_2, \;\;\; L_2 = -k R_1,
	\;\;\; L_3 = -mk, \;\;\; S_1 = \dot R_1, \;\;\; S_2 = \dot R_2 \;\;\; \text{and} \;\;\; S_3 = uk.
	\label{e:EOM-R_3}
	\eeqs
It is clear from (\ref{e: L-and-S}) that $L$ and $S$ do not depend on the coordinate $R_3$. The EOM (\ref{e: EOM-LS}, \ref{e:EOM-R_3}) may be expressed as a system of three second order ODEs for the components of $R(t)$: 
	\beq
	\ddot R_1 = \la k (R_1 \dot R_3 -  m \dot R_2) - k^2 R_1, \;\;
	\ddot R_2 = \la k (R_2 \dot R_3 +  m  \dot R_1) - k^2 R_2 \;\;
	\text{and} \;\;
	\ddot R_3 = \frac{-\la k}{2} ( R_1^2 + R_2^2)_{t}.
	\label{e:EOM-R}
	\eeq
Rajeev and Ranken used conserved quantities to express the solutions to (\ref{e:EOM-R}) in terms of elliptic functions. Here, we examine Hamiltonian and Lagrangian formulations of this model, certain aspects of its classical integrability and explore some properties of its conserved quantities. We also relate this model to the Neumann model and thereby find a new Hamiltonian-Poisson bracket formulation for the latter.

\section{Hamiltonian, Poisson brackets and Lagrangian}
\label{s:Hamiltonian-PB-Lagrangian-RR-model}
\subsection{Hamiltonian and PBs for the RR model}
\label{s:Hamiltonian-mechanical}

This mechanical system with 3 degrees of freedom and phase space $M^6_{S \text{-} L}$ ($\mathbb{R}^6$ with coordinates $L_a, S_a$) can be given a Hamiltonian-Poisson bracket formulation. A Hamiltonian is obtained by a reduction of that of the nilpotent field theory (\ref{e:H-Nilpotent-field-theory}). From (\ref{e:ansatz}), we have $\Tr \dot \phi^2 = \Tr \dot R^2 $ and $\Tr \phi'^2 = \Tr ([K,R] + m K)^2$. Thus the ansatz (\ref{e:ansatz}) has a constant energy density and we define the reduced Hamiltonian to be the energy (\ref{e:H-Nilpotent-field-theory}) per unit length (with dimensions of 1/area):
	\beq
	H = \half \Tr \left[ \left(S-\frac{1}{\la}K\right)^2 +  L^2 \right]
	=  \frac{S_a^2 + L_a^2}{2} + \frac{k}{\la} S_3 + \frac{k^2}{2\la^2} 
	= \half \left[ \dot R_a^2 + k^2 \left(R_1^2 + R_2^2 + m^2 \right) \right].
	\label{e: H-mechanical}
	\eeq
We have multiplied by $\la$ for convenience. PBs among $S$ and $L$ which lead (\ref{e: EOM-LS}) are given by 
	\beq
	\left\{ L_a, L_b \right\}_{\nu} = 0, 
	\quad \left\lbrace S_a, S_b \right\rbrace_{\nu} = \la \epsilon_{abc} L_c
	\quad \text{and} \quad 
	\left\lbrace S_a, L_b \right\rbrace_{\nu} = -\epsilon_{abc} K_c.
	\label{e: PB-SL}
	\eeq
We may view this Poisson algebra as a finite-dimensional version of the nilpotent Lie algebra of currents $I$ and $J$ in (\ref{e:PB-currents}) with $K$ playing the role of the central $\del'$ term. In fact, both are step-3 nilpotent Lie algebras (indicated by $\{ \cdot , \cdot \}_\nu$ in the mechanical model) and we may go from (\ref{e:PB-currents}) to (\ref{e: PB-SL}) via the rough identifications (up to conjugation by $e^{Kx}$):
	\beq
	J_a \to L_a, \quad I_a \to \ov{\la} \left( S_a - \frac{K_a}{\la} \right), \quad  
	\del_{ab} \pdr_x\del(x-y) \to -\eps_{abc} K_c \quad \text{and} \quad \{ \cdot , \cdot \} \to \la \{ \cdot , \cdot \}_{\nu}.
	\eeq
Note that the PBs (\ref{e: PB-SL}) have dimensions of a wave number. They may be expressed as $\{ f, g \}_{\nu} = \scripty{r}^{a b}_0 \pdr_a f \pdr_b g$ where the anti-symmetric Poisson tensor field $\scripty{r}_0 = (0 \: A | A \: B)$ with the $3 \times 3$ blocks $A_{a b} = -\eps_{abc} K_c$ and $B_{a b} = \la \eps_{abc} L_c$.

This Poisson algebra is degenerate: $\scripty{r}_0$ has rank four and its kernel is spanned by the exact 1-forms $d L_3$ and $d\left( S_3 + (\la/k)(L_1^2 + L_2^2)/2 \right)$. The corresponding center of the algebra can be taken to be generated by the Casimirs $m k^2 \equiv \Tr KL $ and ${\mathfrak{c}} k^2 \equiv \Tr \left( (L^2/2) - (KS/\la) \right)$. 
\vspace{.25cm}

{\bf \fl Euclidean PBs:} The $L$-$S$ EOM (\ref{e: EOM-LS}) admit a second Hamiltonian formulation with a non-nilpotent Poisson algebra arising from the reduction of the Euclidean current algebra of the PCM (\ref{e:PB-PCM}). It is straightforward to verify that the PBs
	\beq
	\{ S_a, S_b \}_{\varepsilon} = 0, \quad \{ L_a, L_b \}_{\varepsilon} = - \la \eps_{abc} L_c \quad \text{and} \quad \{ L_a , S_b \}_{\varepsilon} = - \la \eps_{abc} S_c
	\label{e:PB-SL-dual}
	\eeq
along with the Hamiltonian (\ref{e: H-mechanical}) lead to the EOM (\ref{e: EOM-LS}). This Poisson algebra is isomorphic to the Euclidean algebra in 3D (${\mathfrak e}(3)$ or ${\mathfrak{iso}}(3)$) a semi-direct product of the simple $\mathfrak{su}(2)$ Lie algebra generated by the $L_a$ and the abelian algebra of the $S_a$. Furthermore, it is easily verified that $s^2 k^2 \equiv \Tr S^2$ and $h k^2 \equiv \Tr SL$ are Casimirs of this Poisson algebra whose Poisson tensor we denote $\scripty{r}_1$. It follows that the EOM (\ref{e: EOM-LS}) obtained from these PBs are unaltered if we remove the $\Tr S^2$ term from the Hamiltonian (\ref{e: H-mechanical}). The factor $\la$ in the $\{ L_a, S_b\}_{\varepsilon}$ PB is fixed by the EOM while that in the $\{ L_a, L_b \}_{\varepsilon}$ PB is necessary for $h$ to be a Casimir.  
\vspace{.25 cm}

{\fl \bf Formulation in terms of real antisymmetric matrices:} It is sometimes convenient to re-express the $2 \times 2$ anti-hermitian $\mathfrak{su}(2)$ Lie algebra elements $L, S$ and $K$ as $3 \times 3$ real anti-symmetric matrices (more generally we would contract with the structure constants):
	\beq
	\tl L_{k l} = \half \eps_{k l m} L_m \quad \text{with} \quad 
	L_j = \eps_{j k l} \tl L_{k l} \quad \text{and similarly for $\tl S$ and $\tl K$}.
	\eeq 
The EOM (\ref{e: EOM-LS})  and the Hamiltonian (\ref{e: H-mechanical}) become:
	\beq
	{\dot {\tl L}} = -2 [ \tl K , \tl S ], \quad 
	{\dot {\tl S}} = -2 \la [ \tl S, \tl L ] \quad \text{and} \quad 
	H = -\tr \left( \left(\tl S - \tl K/\la \right)^2 + \tl L^2 \right).
	\label{e:EOM-tilde-LS-and-H-RR-real-anti-symm}
	\eeq 
Moreover, the nilpotent $(\nu)$ (\ref{e: PB-SL}) and Euclidean $(\varepsilon)$ (\ref{e:PB-SL-dual}) PBs become
	\beqs
	\{ \tl S_{kl}, \tl S_{pq} \}_{\nu} &=& \frac{\la}{2} \left( \del_{kq} \tl L_{pl} - \del_{pl} \tl L_{kq}  + \del_{ql} \tl L_{kp} - \del_{kp} \tl L_{ql} \right), \cr
	\{ \tl S_{kl}, \tl L_{pq} \}_{\nu} &=& -\half \left( \del_{kq} \tl K_{pl} - \del_{pl} \tl K_{kq} + \del_{ql} \tl K_{kp} - \del_{kp} \tl K_{ql} \right) \quad \text{and} \quad \{ \tl L_{kl} , \tl L_{pq} \}_{\nu} = 0 \quad 
	\label{e:PB-nilpotent-tilde-LS}\\
	\text{and} \quad \{ \tl L_{kl} , \tl L_{pq} \}_{\varepsilon} &=& -\frac{\la}{2} \left( \del_{kq} \tl L_{pl} - \del_{pl} \tl L_{kq}  + \del_{ql} \tl L_{kp} - \del_{kp} \tl L_{ql} \right), \cr
	\{ \tl S_{kl} , \tl L_{pq} \}_{\varepsilon} &=& -\frac{\la}{2} \left( \del_{kq} \tl S_{pl} - \del_{pl} \tl S_{kq} + \del_{ql} \tl S_{kp} - \del_{kp} \tl S_{ql} \right) \quad \text{and} \quad \{ \tl S_{kl}, \tl S_{pq} \}_{\varepsilon} = 0.
	\label{e:PB-semi-diect-tilde-LS}
	\eeqs
Interestingly, we notice that both (\ref{e:PB-nilpotent-tilde-LS}) and (\ref{e:PB-semi-diect-tilde-LS}) display the symmetry $\{ \tl S_{kl}, \tl L_{pq} \} = \{ \tl L_{kl}, \tl S_{pq} \}$. The Hamiltonian (\ref{e:EOM-tilde-LS-and-H-RR-real-anti-symm}) along with either of the PBs (\ref{e:PB-nilpotent-tilde-LS}) or (\ref{e:PB-semi-diect-tilde-LS}) gives the EOM in (\ref{e:EOM-tilde-LS-and-H-RR-real-anti-symm}).

\subsection{Poisson pencil from nilpotent and Euclidean PBs}
\label{s:semi-direct-product-PB}

The Euclidean $\{ \cdot, \cdot \}_{\varepsilon}$ (\ref{e:PB-SL-dual}) and nilpotent $\{ \cdot, \cdot \}_{\nu}$ (\ref{e: PB-SL}) Poisson structures among $L$ and $S$ are compatible and together form a Poisson pencil. In other words, the linear combination
	\beq
	\{ f,g \}_\al = (1- \al) \{ f, g \}_\nu + \al \{ f, g\}_\varepsilon
	\eeq
defines a Poisson bracket for any real $\al$. The linearity, skew-symmetry and derivation properties of the $\al$-bracket follow from those of the individual PBs. As for the Jacobi identity, we first prove it for the coordinate functions $L_a$ and $S_a$. There are only four independent cases:
	\beqs
	\{ \{ S_a , S_b\}_\al, S_c \}_\al + \text{cyclic} &=& -(1 - \al) \la \eps_{abd} \left((1 - \al) \eps_{dce} K_e + \al \la \eps_{dce} S_e \right) + \text{cyclic} = 0, \cr
	\{ \{ L_a, L_b\}_\al , L_c \}_\al + \text{cyclic} &=& \al^2 \la^2 \eps_{abd} \eps_{dce} L_e + \text{cyclic} = 0, \cr
	\{ \{ S_a, S_b \}_\al, L_c \}_\al + \text{cyclic} &=& -(1- \al) \al \la^2 \eps_{abd} \eps_{dce} L_e + \text{cyclic} = 0 \quad  \text{and} \cr
	\{ \{ L_a, L_b\}_\al , S_c \}_\al + \text{cyclic} &=& \al \la \eps_{abd} \left( (1- \al)\eps_{dce} K_e + \al \la \eps_{dce} S_e \right) + \text{cyclic} = 0.
	\label{e:Jacobi-identity}
	\eeqs 
The Jacobi identity for the $\al$-bracket for linear functions of $L$ and $S$ follows from (\ref{e:Jacobi-identity}). For more general functions of $L$ and $S$, it follows by applying the Leibniz rule ($\xi_i = (L_{1,2,3},S_{1,2,3})$):
	\beq
	\{ \{ f, g\}_\al, h \}_\al + \text{cyclic} = \dd{f}{\xi_i} \dd{g}{\xi_j} \dd{h}{\xi_k} \left( \{ \{ \xi_i, \xi_j \}_\al , \xi_k \}_\al + \text{cyclic} \right) = 0.
	\eeq 

As noted, both the nilpotent and Euclidean PBs are degenerate: ${\mathfrak{c}}$ and $m$ are Casimirs of $\{ \cdot , \cdot \}_\nu$ while those of $\{ \cdot, \cdot \}_\veps$ are $s^2$ and $h$. In fact, the Poisson tensor $\scripty{r}_\al = (1- \al) \scripty{r}_{0} + \al \scripty{r}_{1}$ is degenerate for any $\alpha$ and has rank 4. Its independent Casimirs may be chosen as $(1- \al)( m/ \la ) + \al h$ and $(1 - \al) {\mathfrak{c}} - \al  s^2/2$, whose exterior derivatives span the kernel of $\scripty{r}_\al$. The $\nu$ and $\varepsilon$ PBs become non-degenerate upon reducing the 6D phase space to the 4D level sets of the corresponding Casimirs. Since the Casimirs are different, the resulting symplectic leaves are different, as are the corresponding EOM. Thus these two PBs do not directly lead to a bi-Hamiltonian formulation.

\subsection{Darboux coordinates and Lagrangian from Hamiltonian}
\label{s:Darboux-coordinates-M6}
	
Though they are convenient, the $S$ and $L$ variables are non-canonical generators of the nilpotent degenerate Poisson algebra (\ref{e: PB-SL}). Moreover, they lack information  about the coordinate $R_3$. It is natural to seek canonical coordinates that contain information on all six generalized coordinates and velocities $(R_a, \dot R_a)$ (see (\ref{e:currents-I-J})). Such Darboux coordinates will also facilitate a passage from Hamiltonian to Lagrangian. Unfortunately, as discussed below, the naive reduction of (\ref{e:Lagrangian-scalar-field}) does not yield a Lagrangian for the EOM (\ref{e:EOM-R}). 

It turns out that momenta conjugate to the coordinates $R_a$ may be chosen as (see (\ref{e:EOM-R_3}))
	\beqs
	kP_1 &=& S_1 + \frac{\la}{2} m L_1 = \dot R_1 + \frac{\la}{2}m k R_2, 
	\quad kP_2 = S_2 + \frac{\la}{2} m L_2 = \dot R_2 - \frac{\la}{2}m k R_1  \quad \text{and} \cr
	kP_3 &=& \frac{k\la}{2} ( 2 {\mathfrak{c}} - m^2 ) + \frac{k}{\la} = S_3 + \frac{k}{\la} + \frac{\la}{2k} \left( L_1^2 + L_2^2 \right) = \dot R_3 + \frac{\la k}{2}(R_1^2 + R_2^2).
	\label{e:canonical-momenta-P_a}
	\eeqs
We obtained them from the nilpotent algebra (\ref{e: PB-SL}) by requiring the canonical PB relations
	\beq
	\{ R_a, R_b \} = 0, 
	\quad \{ P_a, P_b \} = 0 
	\quad \text{and} \quad \{ R_a, k P_b \} = \delta_{a b} 
	\quad \text{for} \quad a,b = 1,2,3.
	\label{e:Canonical-PB-RP}
	\eeq
Note that $R_a$ cannot be treated as coordinates for the Euclidean PBs (\ref{e:PB-SL-dual}), since $\{ R_1, R_2 \} = (1/k^2) \{ L_1, L_2 \}_{\varepsilon} \neq 0$. Darboux coordinates associated to the Euclidean PBs, may be analogously obtained from the coordinates $Q$ in the wave ansatz for the mechanical reduction of the principal chiral field $g = e^{\la s K x} Q(t) e^{-Kx}$ given in Table I of \cite{R-R}. 

Since $R_3$ does not appear in the Hamiltonian (\ref{e: H-mechanical}) (regarded as a function of $(S, L)$ or $(R, \dot R)$), we have taken the momenta in (\ref{e:canonical-momenta-P_a}) to be independent of $R_3$ so that it will be cyclic in the Lagrangian as well. However, the above formulae for $P_a$ are not uniquely determined. For instance, the PBs (\ref{e:Canonical-PB-RP}) are unaffected if we add to $P_a$ any function of the Casimirs $({\mathfrak{c}}, m)$ as also certain functions of the coordinates (see below for an example). In fact, we have used this freedom to pick $P_3$ to be a convenient function of the Casimirs. Moreover, $\{ R_3, k P_3 \} = 1$ is a new postulate, it is not a consequence of the $S$-$L$ Poisson algebra.
  
The Hamiltonian (\ref{e: H-mechanical}) can be expressed in terms of the $R$'s and $P$'s:
	\beq
	 \frac{H}{k^2} = \sum_{a=1}^3 \frac{P_a^2}{2} + \frac{\la m}{2} \left( R_1 P_2 - R_2 P_1 \right) + \frac{\la^2}{8} \left( R_1^2 + R_2^2 \right) \left[ R_1^2 + R_2^2 + m^2 - \frac{4}{\la} \left( P_3 - \frac{1}{\la} \right) \right] + \frac{m^2}{2}.
	\label{e:Hamiltonian-mech-R-P}
	\eeq
The EOM (\ref{e: EOM-LS}), (\ref{e:EOM-R_3}) follow from (\ref{e:Hamiltonian-mech-R-P}) and the PBs (\ref{e:Canonical-PB-RP}). Thus $R_a$ and $kP_b$ are Darboux coordinates on the 6D phase space $M^6_{R \text{-} P} \cong \mathbb{R}^6$. Note that the previously introduced phase space $M^6_{S \text{-} L}$ is different from $M^6_{R \text{-} P}$, though they share a 5D submanifold in common parameterized by $(L_{1,2}, S_{1,2,3})$ or $(R_{1,2}, P_{1,2,3})$. $M^6_{S \text{-} L}$ includes the constant parameter $L_3 = - m k$ as its sixth coordinate but lacks information on $R_3$ which is the `extra' coordinate in $M^6_{R \text{-} P}$.

\vspace{.25cm}

{\fl \bf Lagrangian for the RR model:} A Lagrangian $L_{\rm mech}(R,\dot R)$ for our system may now be obtained via a Legendre transform by extremizing $kP_a \dot R_a - H$ with respect to all the components of $kP$:
	\beq
	L_{\rm mech} = \half \left[ \sum_{a=1}^3 \dot R_a^2 - \la m k \left( R_1 \dot{R_2} - R_2 \dot{R_1} \right) +  k \left( R_1^2 + R_2^2 \right) ( \la \dot{R_3} -  k) -  m^2 k^2 \right].
	\label{e:Lagrangian-Mech}
	\eeq 
$R_3$ is a cyclic coordinate leading to the conservation of $kP_3$. However $L_{\rm mech}$ does not admit an invariant form as the trace of  a polynomial in $R$ and $\dot R$. Such a form may be obtained by subtracting the time derivative of $(\la k/6) \left( R_3 (R_1^2 + R_2^2) \right)$ from $L_{\rm mech}$ to get: 
	\beqs
	L_{\rm mech}' &=& \Tr \left( \frac{\dot R^2}{2} - \half  ([K, R] + mK)^2 + \frac{\la}{2} R [\dot R, mK] + \frac{\la}{3} R \left[\dot R , [ K, R ] \right] \right) \cr
	 &=& \half \Tr \left( \left(S - \frac{K}{\la} \right)^2 - L^2 + \la R \left[ S - \frac{K}{\la} , L \right] - \frac{\la}{3} R \left[ S - \frac{K}{\la} , [K, R] \right] \right).
	\label{e:Lagrangian-mech}
	\eeqs 
The price to pay for this invariant form is that $R_3$ is no longer cyclic, so that the conservation of $P_3$ is not manifest. The Lagrangian $L_{\rm mech}'$ may also be obtained directly from the Hamiltonian (\ref{e:Hamiltonian-mech-R-P}) if we choose as conjugate momenta $k\Pi_a$ instead of the $kP_a$ of (\ref{e:canonical-momenta-P_a}):
	\beq
	\Pi_1 = P_1 - \frac{\la}{3} R_1 R_3,\quad  
	\Pi_2 = P_2 - \frac{\la}{3} R_2 R_3 \quad \text{and} \quad
	\Pi_3 = P_3 - \frac{\la}{6}(R_1^2 + R_2^2). 
	\eeq
Interestingly, while both $L_{\rm mech}$ and $L_{\rm mech}'$ give the correct EOM (\ref{e:EOM-R}), unlike with the Hamiltonian, the naive reduction $L_{\rm naive}$ of the field theoretic Lagrangian (\ref{e:Lagrangian-scalar-field}) does not. This discrepancy was unfortunately overlooked in Eq.~(3.7) of \cite{R-R}. Indeed $L_{\rm naive}$ differs from $L_{\rm mech}'$ by a term which is {\it not} a time derivative: 
	\beq
	L_{\rm naive} = L_{\rm mech}' + \frac{\la m}{6} \Tr  K \:[ \dot R, R].
	\eeq
To see this, we put the ansatz (\ref{e:ansatz}) for $\phi$ in the nilpotent field theory Lagrangian (\ref{e:Lagrangian-scalar-field}) and use
	\beqs
	\Tr \dot \phi^2 &=& \Tr \dot R^2 , \quad \Tr \phi'^2 = \Tr ([K,R] + m K)^2 \quad \text{and} \cr
	\Tr \phi [\dot \phi, \phi'] &=& \Tr R \left[\dot R ,[K,R]+ mK \right] + \frac{m x k^2}{2} \DD{}{t} (R_1^2 + R_2^2) 
	\eeqs
to get the naively reduced Lagrangian  
	\beq
	L_{\rm naive} = \Tr \left( \half \dot R^2 + \frac{\la}{3} R \left[ \dot R , \left[ K , R \right] + mK \right] - \half (\left[K, R\right] + mK)^2\right).
	\eeq
In obtaining $L_{\rm naive}$ we have ignored an $x$-dependent term as it is a total time derivative, a factor of the length of space and multiplied through by $\la$. As mentioned earlier, $L_{\rm naive}$ does {\it not} give the correct EOM for $R_1$ and $R_2$ nor does it lead to the PBs among $L$ and $S$ (\ref{e: PB-SL}) if we postulate canonical PBs  among $R_a$ and their conjugate momenta. However the Legendre transforms of $L_{\rm mech}, L_{\rm mech}'$ and $L_{\rm naive}$ all give the same Hamiltonian (\ref{e: H-mechanical}). 

One may wonder how it could happen that  the naive reduction of the scalar field gives a suitable Hamiltonian  but not a suitable Lagrangian for the mechanical system. The point is that while a Lagrangian encodes the EOM, a
Hamiltonian by itself does not. It needs to be supplemented with PBs. In the present case, while we used a naive reduction of the scalar field Hamiltonian as the Hamiltonian for the RR model, the relevant PBs ((\ref{e: PB-SL}) and (\ref{e:Canonical-PB-RP})) are not a simple reduction of those of the field theory ((\ref{e:PB-currents}) and (\ref{e: Canonical-PB-field})). Thus, it is not surprising that the naive reduction of the scalar field Lagrangian does not furnish a suitable Lagrangian for the mechanical system. This possibility was overlooked in \cite{R-R} where the former was proposed as a Lagrangian for the RR model.

\section{Lax pairs, $r$-matrices and conserved quantities}
\subsection{Lax Pairs and $r$-matrices}
\label{s:Lax-pair-and-r-matrix}

The EOM (\ref{e: EOM-LS}) admit a Lax pair $(A, B)$ with complex spectral parameter $\zeta$. In other words, if we choose
 	\beq
	A(\zeta) = -K\zeta^2 + L\zeta + \frac{S}{\la} 
	\quad \text{and} \quad	
	B(\zeta) = \frac{S}{\zeta},
	\label{e:Lax-pair}
	\eeq
then the Lax equation $\dot A = [B, A]$ at orders $\zeta^1$ and $\zeta^0$ are equivalent to (\ref{e: EOM-LS}). The Lax equation implies that $\Tr A^n(\zeta)$ is a conserved quantity for all $\zeta$ and every $n = 1,2,3 \ldots$. To arrive at this Lax pair we notice that $\dot A = [B,A]$ can lead to (\ref{e: EOM-LS}) if $L$ and $S$ appear linearly in $A$ as coefficients of different powers of $\zeta$. The coefficients have been chosen to ensure that the fundamental PBs (FPBs) between matrix elements of $A$ can be expressed as the commutator with a non-dynamical $r$-matrix proportional to the permutation operator. In fact, the FPBs with respect to the nilpotent PBs (\ref{e: PB-SL}) are given by
	\beqs
	\{ A(\zeta) \stackrel{\otimes}{,} A(\zeta') \}_\nu
	&=& -\frac{1}{4 \la}\left(\epsilon_{abc}L_c -\epsilon_{abc} K_c \left(\zeta + \zeta'\right) \right) \sig_a \otimes \sig_b \cr
	&=& \frac{i}{2 \la}\left(L_3 - \left(\zeta + \zeta'\right) K_3  \right)  \left(\sigma_- \otimes \sigma_+ - \sigma_+ \otimes \sigma_- \right) \cr 
	&& + \frac{1}{4 \la} \sum_\pm \left( {L_2 \pm i L_1} \right)\left(\sigma_\pm \otimes \sigma_3 - \sigma_3 \otimes \sigma_\pm \right).
	\label{e:FPB-nilpotent}
	\eeqs
Here, $\sig_\pm = (\sig_1 \pm i \sig_2)/2$. These FPBs can be expressed as a commutator
	\beqs
	\left\lbrace A(\zeta) \stackrel{\otimes}{,} A(\zeta') \right\rbrace_\nu &=& \left[ r(\zeta, \zeta') , A(\zeta)\otimes I + I \otimes A(\zeta')\right]  \quad \text{where}\cr
	r(\zeta, \zeta') &=& - \frac{P}{2\la (\zeta - \zeta')} 
	\quad \text{with} \quad 
	P = \half \left(I + \sum_{a=1}^3 \sigma_a \otimes \sigma_a \right).
	\label{e:r-matrix-nilpotent}
	\eeqs
To obtain this $r$-matrix we used the following identities among Pauli matrices:
	\beqs
	\sigma_- \otimes \sigma_+ - \sigma_+ \otimes \sigma_- &=& \half [ P, \sigma_3 \otimes I ] = -\half [ P , I \otimes \sigma_3 ]
	\quad \text{and} \cr
	\sigma_\pm \otimes \sigma_3 - \sigma_3 \otimes \sigma_\pm &=& \pm\left[ P , \sigma_\pm \otimes I \right] = \mp \left[ P, I \otimes \sigma_\pm \right].
	\eeqs
We may now motivate the particular choice of Lax matrix $A$ (\ref{e:Lax-pair}). The  nilpotent $S$-$L$ PBs (\ref{e: PB-SL}) do not involve $S$, so the PBs between matrix elements of $A$ are also independent of $S$. Since $P(A \otimes B) =  (B \otimes A) P$, the commutator $\left[P, A \otimes I + I \otimes A \right] = 0$ if $A$ is independent of $\zeta$. Thus for $r \propto P$, $S$ can only appear as the coefficient of $\zeta^0$ in $A$.

The same commutator form of the FPBs (\ref{e:r-matrix-nilpotent}) hold for the Euclidean PBs (\ref{e:PB-SL-dual})  if we use
	\beq
	r_{\varepsilon}(\zeta, \zeta') = \la^2 r(\zeta, \zeta') =  -\frac{\la P}{2 (\zeta - \zeta')},
	\label{e:r-matrix-semi-direct}
	\eeq
provided we define a new Lax matrix $A_\varepsilon = A/\zeta^2$. The EOM for $S$ and $L$ are then equivalent to the Lax equation $\dot A_{\varepsilon} = [B, A_{\varepsilon}]$ at order $\zeta^{-2}$ and $\zeta^{-1}$. In this case, the FPBs are
	\beq	
	\{ A_{\varepsilon}(\zeta) \stackrel{\otimes}{,} A_{\varepsilon}(\zeta') \}_\varepsilon
	= \frac{1}{4 \zeta \zeta'} \left(\la \epsilon_{abc}L_c + \left(\ov{\zeta} + \ov{\zeta'} \right) \epsilon_{abc} S_c \right) \sig_a \otimes \sig_b. 
	\label{e:FPB-semi-direct}
	\eeq

\subsection{Conserved quantities in involution for the RR model}
\label{s:conserved-quantities}

Eq.~(\ref{e:r-matrix-nilpotent}) for the FPBs implies that the conserved quantities $\Tr A^n(\zeta)$ are in involution: 
	\beq
	\left\{\Tr A^m(\zeta) \stackrel{\otimes}{,} \Tr A^n(\zeta') \right\} = mn \Tr \left[ r(\zeta,\zeta') , A^m(\zeta)\otimes A^{n-1}(\zeta') + A^{m-1}(\zeta)\otimes A^{n}(\zeta') \right] = 0
	\eeq
for $m,n = 1,2,3 \ldots$. Each coefficient of the $2n^{\rm th}$ degree polynomial $\Tr A^n(\zeta)$ furnishes a conserved quantity in involution with the others. However, they cannot all be independent as the model has only 3 degrees of freedom. For instance, $\Tr A(\zeta) \equiv 0$ but 
	\beq
	\Tr A^2(\zeta) = \zeta^4 \: K_a K_a - 2 \zeta^3 \: L_a K_a + 2 \zeta^2 \left( \frac{L_a L_a}{2} - \frac{S_a K_a}{\la} \right) + \frac{2\zeta}{\la} \: S_a L_a + \frac{1}{\la^2} S_a S_a.
	\eeq
In this case, the coefficients give four conserved quantities in involution:
	\beqs
	s^2 k^2 &=& \Tr S^2, \quad
	h k^2 = \Tr SL, \quad
	m k^2 = \Tr KL = - k L_3 \cr
	\text{and} \quad
	{\mathfrak{c}} k^2 &=& \Tr \left(\frac{L^2}{2} - \frac{1}{\la}KS \right) = \half L_a L_a + \frac{k}{\la} S_3.
	\label{e:conserved-quantities}
	\eeqs
Factors of $k^2$ have been introduced so that ${\mathfrak{c}}$, $m$, $h$ and $s^2$ (whose positive square-root we denote by $s$) are dimensionless. In \cite{R-R}, $h$ and ${\mathfrak{c}}$ were named $C_1$ and $C_2$. ${\mathfrak{c}}$ and $m$ may be shown to be Casimirs of the nilpotent Poisson algebra (\ref{e: PB-SL}). The value of the Casimir $L_3$ is written as $-m$ in units of $k$ by analogy with the eigenvalue of the angular momentum component $L_z$ in units of $\hbar$. The conserved quantity $\Tr SL$ is called $h$ for helicity by analogy with other such projections. The Hamiltonian (\ref{e: H-mechanical}) can be expressed in terms of $s^2$ and $\mathfrak{c}$:
	\beq
	H = k^2 \left(\half s^2+ {\mathfrak{c}} + \ov{2\la^2} \right).
	\label{e:Hamiltonian-s}
	\eeq
It will be useful to introduce the 4D space of conserved quantities $\cal Q$ with coordinates $\mathfrak{c}$, $s$, $m$ and $h$ which together define a many-to-one map from $M^6_{S \text{-} L}$ to $\cal Q$. The inverse images of points in $\cal Q$ under this map define common level sets of conserved quantities in $M^6_{S \text{-} L}$. By assigning arbitrary real values to the Casimirs $\mathfrak{c}$ and $m$ we may go from the 6D $S$-$L$ phase space to its non-degenerate $4$D symplectic leaves $M^4_{{\mathfrak{c}} m}$ given by their common level sets. For the reduced dynamics on $M^4_{{\mathfrak{c}} m}$, $s^2$ (or $H$) and $h$ define two conserved quantities in involution. 

\begin{figure}[h]
	\centering
		\begin{subfigure}[b]{5cm}
		\includegraphics[width=5cm]{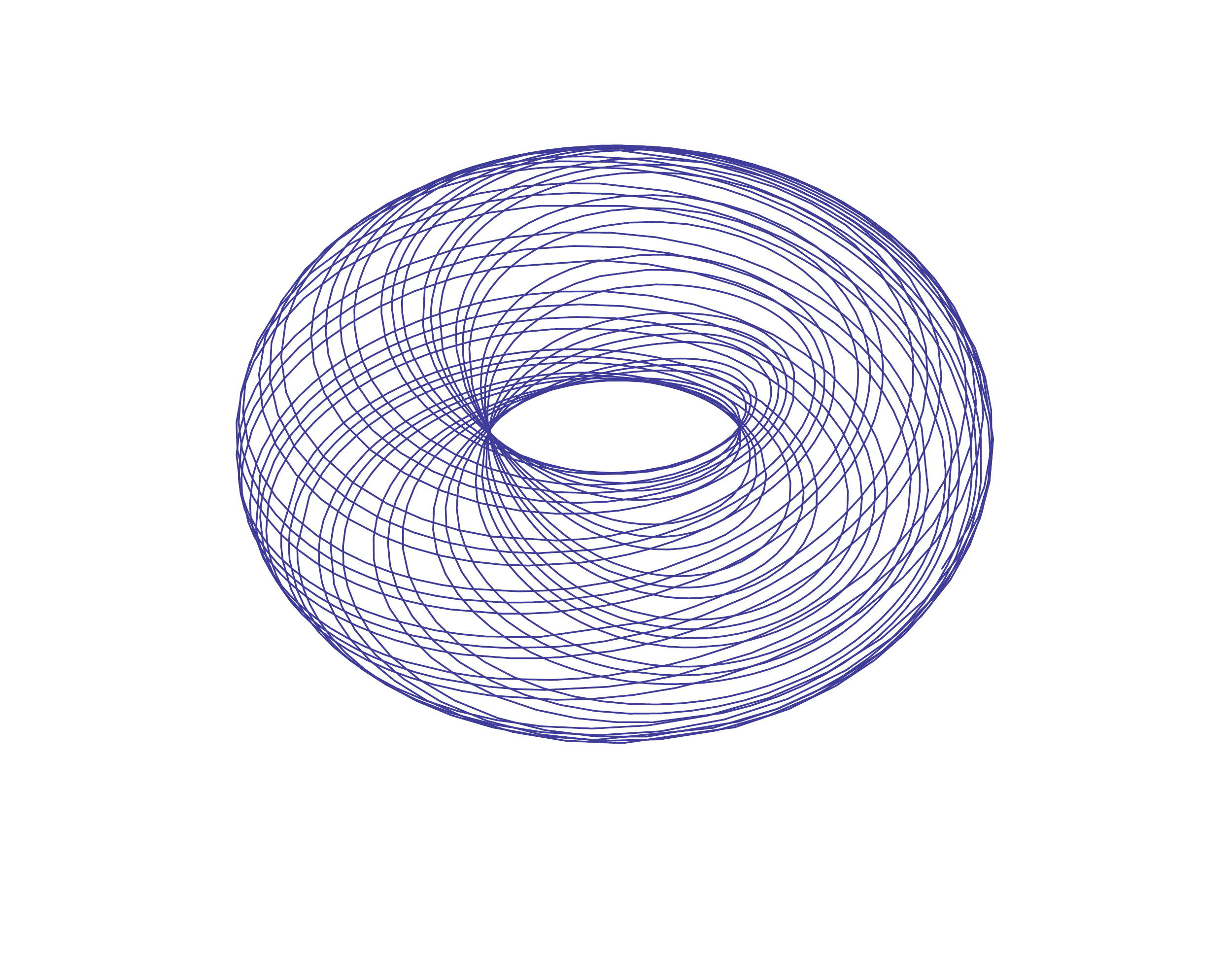}
		\end{subfigure}
	\caption{\footnotesize A trajectory  with initial conditions $\tht(0)= 0.1$ and $\phi(0) = 0.2$ plotted for $0 \leq t \leq 200/k$ on a generic common level set of the conserved quantities ${\mathfrak{c}}, m, s$ and $h$. The common level set is a 2-torus parameterized by the polar and azimuthal angles $\tht$ and $\phi$ and has been plotted for the values ${\mathfrak{c}} = 1/2, h = 0, m = s = 1$ with $k = \la = 1$. It is plausible that the trajectory is quasi-periodic and dense on the torus so that any additional conserved quantity would have to be a constant.}
	\label{f:theta-phi-dynamics-3D}
	\end{figure}

The independence of ${\mathfrak{c}}, m , h$ and $s$ is discussed in \S \ref{s:independence}. However, higher powers of $A$ do not lead to new conserved quantities. $\Tr A^3 \equiv 0$ since $\Tr (t_a t_b t_c) = \half \epsilon_{abc}$ for $t_a = \sig_a/2i$. The same applies to other odd powers. On the other hand, the expression for $A^4(\zeta)$ given in Appendix \ref{a:A^4}, along with the identity $\Tr (t_a t_b t_c t_d) = -\frac{1}{4} (\del_{ab} \del_{cd} - \del_{ac} \del_{bd} + \del_{ad} \del_{bc})$ gives
	\beqs
	\ov{k^4} \Tr A^4(\zeta) &=& -\ov 4 s^4 - h s^2 \zeta - \left( \frac{{\mathfrak{c}}s^2 + h^2 }{\la^2} \right) \zeta^2 -  \left(\frac{2 h {\mathfrak{c}}}{\la} - \frac{ms^2}{\la^2} \right) \zeta^3 -  \left( {\mathfrak{c}}^2  + \frac{s^2 }{\la^2} - \frac{2}{\la} m h  \right) \zeta^4 \cr && +  \left( m{\mathfrak{c}} - \ov \la h \right) \zeta^5 - \left( {\mathfrak{c}} + \half m + 2 m^2 \right)\zeta^6 + \ov 4 m \zeta^7 - \ov 4  \zeta^8.
	\label{e:trace-A4}
	\eeqs
Evidently, the coefficients of various powers of $\zeta$ are functions of the known conserved quantities (\ref{e:conserved-quantities}). It is possible to show that the higher powers $\Tr A^6, \Tr A^8 , \ldots$ also cannot yield new conserved quantities by examining the dynamics on the common level sets of the known conserved quantities. In fact, we find that a generic trajectory (obtained by solving (\ref{e:theta-phi-dynamics})) on a generic common level set of all four conserved quantities is dense (see Fig. \ref{f:theta-phi-dynamics-3D} for an example). Thus, any additional conserved quantity would have to be constant almost everywhere and cannot be independent of the known ones.

\vspace{.25cm}

{\fl \bf Canonical vector fields on $M^6_{S \text{-} L}$:} On the phase space, the canonical vector fields ($V_f^a = \scripty{r}_0^{a b} \pdr_b f$) associated to conserved quantities, follow from the Poisson tensor of \S \ref{s:Hamiltonian-mechanical}. They vanish for the Casimirs ($V_{\mathfrak{c}} = V_{m} = 0$) while for helicity and the Hamiltonian $(H = E k^2)$,
	\beqs
	k V_h &=&  L_2 \pdr_{L_1} -  L_1 \pdr_{L_2} + S_2 \pdr_{S_1} -  S_1 \pdr_{S_2}    \quad \text{and}  \cr
	k V_E &=& S_2 \pdr_{L_1} - S_1 \pdr_{L_2} + \frac{\la}{k} \left[(S_2 L_3 - L_2 S_3) \pdr_{S_1} + (S_3 L_1 - S_1 L_3) \pdr_{S_2} + (S_1 L_2 - S_2 L_1) \pdr_{S_3} \right]. \qquad
	\label{e:Hamiltonian-V-F}
	\eeqs
The coefficient of each of the coordinate vector fields in $V_E$ gives the time derivative of the corresponding coordinate (upto a factor of $k^2$) and leads to the EOM (\ref{e:EOM-LS-explicit}). These vector fields commute, since $[ V_E, V_h ] = - V_{\{E, h \}}$.

\vspace{.25cm}

{\fl \bf Conserved quantities for the Euclidean Poisson algebra:} As noted, the same Hamiltonian (\ref{e: H-mechanical}) with the $\{ \cdot, \cdot \}_\varepsilon$ PBs leads to the $S$-$L$ EOM (\ref{e: EOM-LS}). Moreover, it can be shown that ${\mathfrak{c}}, m, s$ and $h$ (\ref{e:conserved-quantities}) continue to be in involution with respect to $\{ \cdot, \cdot \}_\varepsilon$ and to commute with $H$. Interestingly, the Casimirs (${\mathfrak{c}}, m$) and non-Casimir conserved quantities $(s^2, h)$ exchange roles in going from the nilpotent to the Euclidean Poisson algebras.

\vspace{.25cm}

{\fl \bf Simplification of EOM  using conserved quantities:} Using the conserved quantities we may show that $\dot{u}, \dot{\tht}$ and $\dot{\phi}$ are functions of $u = S_3/k$ alone. Indeed, using (\ref{e: PB-SL}) and (\ref{e:L-S-polar}) we get
	\beqs
	\dot{u}^2 &=& \frac{\dot S_3^2}{k^2} = \la^2  k^2 \rho^2 r^2 \sin^2(\theta - \phi), \qquad
	\dot \tht = \frac{L_1 \dot L_2 - \dot L_1 L_2}{L_1^2 + L_2^2}  = - \frac{k \rho}{r} \cos(\tht - \phi) \cr
\text{and} \quad 
	\dot \phi &=& \frac{S_1 \dot S_2 - \dot S_1 S_2}{S_1^2 + S_2^2} = k m \la + k \la   \frac{r u}{\rho} \cos(\tht - \phi).
	\label{e:theta-phi-u-EOM} 
	\eeqs
Now $r,\rho$ and $\tht - \phi$ may be expressed as functions of $u$ and the conserved quantities. In fact,
	\beq
	\rho^2 = s^2 -u^2, \quad  r^2 = 2{\mathfrak{c}} - m^2 - \frac{2u}{\la} \quad  \text{and} \quad h = \frac{\Tr S L}{k^2} = -m u + r \rho  \cos(\tht - \phi).
	\label{e:relation-theta-phi-u}
	\eeq
Thus we arrive at
	\beq
	\dot{u}^2 = \la^2  k^2 \left[ (s^2 - u^2)\left(2{\mathfrak{c}} -m^2 - \frac{2u}{\la}\right) - (h + mu)^2\right] = 2 \la k^2 \chi(u),
	\label{e:EOM-u}
	\eeq
	\beq
	\dot{\theta} = -k\left(\frac{h + m u}{2{\mathfrak{c}} - m^2 - \frac{2u}{\la}}\right) \quad \text{and} \quad
	\dot{\phi} = k m \la + k \la u \left(\frac{h + mu}{s^2 - u^2}\right).
	\label{e:theta-phi-dynamics}
	\eeq
Moreover, the formula for $h$ in (\ref{e:relation-theta-phi-u}) gives a relation among $u, \tht$ and $\phi$ for given values of conserved quantities. Thus, starting from the 6D $S$-$L$ phase space and using the four conservation laws, we have reduced the EOM to a pair of ODEs on the common level set of conserved quantities. For generic values of the conserved quantities, the latter is an invariant torus parameterized, say, by $\tht$ and $\phi$. Furthermore, $\dot u^2$ is proportional to the cubic $\chi(u)$ and may be solved in terms of the $\wp$ function while $\tht$ is expressible in terms of the Weierstrass $\zeta$ and $\sigma$ functions as shown in Ref. \cite{R-R}.

\subsection{Symmetries and associated  canonical transformations}
\label{s:Noether-symmetries-CT}

Here, we identify the Noether symmetries and canonical transformations (CT) generated by the conserved quantities. The constant $m = -L_3/k$ commutes (relative to $\{ \cdot, \cdot \}_{\nu}$) with all observables and acts trivially on the coordinates $R_a$ and momenta $P_b$ of the mechanical system.

The infinitesimal CT $R_3 \to R_3 + \varepsilon$ corresponding to the cyclic coordinate in $L_{\rm mech}$ (\ref{e:Lagrangian-Mech}) is generated by $ (\varepsilon \la k/2) (2 {\mathfrak{c}} - m^2) = \varepsilon k (P_3 - 1/ \la)$ (\ref{e:canonical-momenta-P_a}). $L_{\rm mech}$ is also invariant under infinitesimal rotations in the $R_1$-$R_2$ plane. This corresponds to the infinitesimal CT
	\beq
	\del R_a = \varepsilon \eps_{ab} R_b, \quad
	\del P_a = \varepsilon \eps_{ab} P_b \quad \text{for} \quad a,b = 1,2 \quad \text{and} \quad
	\del R_3 = \del P_3 = 0,
	\eeq
with generator (Noether charge) $\varepsilon k \left[ h + (\la m/ 2) (2 {\mathfrak{c}} - m^2 ) \right]$. The additive constants involving $m$ may of course be dropped from these generators. Thus, while $P_3$ (or equivalently $\mathfrak{c}$) generates translations in $R_3$, $h$ (up to addition of a multiple of $P_3$) generates rotations in the $R_1$-$R_2$ plane. In addition to these two point-symmetries, the Hamiltonian (\ref{e:Hamiltonian-mech-R-P}) is also invariant under an infinitesimal CT that mixes coordinates and momenta:
	\beqs
	\del R_{a} &=& 2 \varepsilon P_a, 
	\quad \del P_a = \varepsilon \la^2 \left[\frac{2}{\la}\left( P_3 - \frac{1}{\la} \right) - (R_1^2 + R_2^2) - \frac{m^2}{2} \right] R_a \quad
	\text{for} \quad a = 1,2
	\cr \text{while} \quad \del R_3 &=& \varepsilon \left[2 P_3 - \la (R_1^2 + R_2^2)\right] \quad \text{and} \quad \del P_3 = 0.
	\eeqs 
This CT is generated by the conserved quantity 
	\beq
	\varepsilon k \left[ s^2 + 2 {\mathfrak{c}} + \la m \left( h + \left(\frac{\la m}{2} \right)(2 {\mathfrak{c}} - m^2 )\right)  \right]
	\label{e:generator-momenta}
	\eeq
which differs from $s^2$ by terms involving $h$ and $\mathfrak{c}$ which serve to simplify the CT by removing an infinitesimal rotation in the $R_1$-$R_2$ plane as well as a constant shift in $R_3$. Here, upto Casimirs, (\ref{e:generator-momenta}) is related to the Hamiltonian via $s^2 + 2 {\mathfrak{c}} = (1/k^2)(2 H - k^2/\la^2)$. 

The above assertions follow from using the canonical PBs, $\{ R_a , k P_b \} = \del_{ab}$ to compute the changes $\del R_a = \{ R_a , Q \}$ etc., generated by the three conserved quantities $Q$ expressed as:
	\beqs
	h &=& P_1 R_2 - P_2 R_1 - m P_3, \qquad
	{\mathfrak{c}} = \frac{1}{\la} \left( P_3 - \frac{1}{\la} \right) + \frac{m^2}{2} \quad \text{and} 
	\cr
	s^2 &=& \sum_{a=1}^3 P_a^2  
	+ \la m \eps_{ab} R_a P_b - \frac{2}{\la} P_3 + \frac{\la^2}{4} \left(R_1^2 + R_2^2 \right) \left[ R_1^2 + R_2^2 - \frac{4}{\la} \left( P_3 - \frac{1}{\la} \right) + m^2 \right] + \frac{1}{\la^2}. \qquad
	\eeqs

\subsection{Relation of conserved quantities to Noether charges of the field theory}
\label{s:conserved-qtys-Noether-charges}

Here we show that three out of four combinations of conserved quantities ($P_3, h - m/\la$ and  $H$) are reductions of scalar field Noether charges, corresponding to symmetries under translations of $\phi$, $x$ and $t$. The fourth conserved quantity $L_3 = -mk$ arose as a parameter in (\ref{e:ansatz}) and is not the reduction of any Noether charge. By contrast, the charge corresponding to internal rotations of $\phi$ does not reduce to a conserved quantity of the RR model. 




Under the shift symmetry $\phi \to \phi + \eta $ of (\ref{e:Non-linear-wave-equation}), the PBs (\ref{e: Canonical-PB-field}) preserve their canonical form as $\del \pi = (1/3) [\eta , \phi']$ commutes with $\phi$. This leads to the conserved Noether density and current
	\beq
	j_t =  \Tr  \eta \left( \frac{\dot \phi}{\la}  - \frac{ [ \phi , \phi']}{2}\right) 
	\quad \text{and} \quad
	j_x = \Tr \eta \left( -\frac{\phi'}{\la} + \frac{ [ \phi,  \dot \phi ]}{2}\right).
	\eeq
The conservation law $\pdr_t j_t + \pdr_x j_x = 0$ is equivalent to (\ref{e:Non-linear-wave-equation}) \cite{C-Z}. Taking $\eta \propto \la$, all matrix elements of $Q^{\rm s} = \int \left( \dot \phi - (\la/2) [ \phi , \phi']\right) \: dx$ are conserved. To obtain $P_3$ (\ref{e:canonical-momenta-P_a}) as a reduction of $Q^{\rm s}$ we insert the ansatz (\ref{e:ansatz}) to get 
	\beq
	Q^{\rm s} = \int e^{Kx} \tl Q^{\rm s} e^{-Kx} \: dx \quad \text{where} \quad  \tl Q^{\rm s} = \dot R - \frac{\la}{2} [R, [ K, R] + mK].
	\eeq	
Expanding $\tl Q^{\rm s} = \tilde{Q}^{\rm s}_a t_a$ and using the Baker-Campbell-Hausdorff formula we may express
	\beq
	Q^{\rm s} = \int \left( \cos kx \sig_2 - \sin kx \sig_1 \right) \frac{\tilde{Q}^{\rm s}_1}{2i} \: dx + \int \left( \cos kx \sig_1 + \sin kx \sig_2 \right) \frac{\tilde{Q}^{\rm s}_2}{2i} \: dx  + \int \tilde{Q}^{\rm s}_3 \frac{\sig_3}{2i} \:dx.
	\eeq
The first two terms vanish while $\tl{Q}^{\rm s}_3 = P_3$ so that $Q^{\rm s} = l P_3 t_3$, where $l$ is the spatial length.



The density $({\cal P} = \Tr \dot{\phi} \phi'/\la)$ and current $(-{\cal E} = -  (1/2\la) \Tr ( \dot \phi^2 + \phi'^2))$ (\ref{e:H-Nilpotent-field-theory}) corresponding to the symmetry $x \to x + \eps$ of (\ref{e:Non-linear-wave-equation}) satisfy $\pdr_t {\cal P} - \pdr_x {\cal E} = 0$ or $\Tr \left( \ddot \phi - \phi'' \right) \phi' = 0$. The conserved momentum $P = \Tr \int I J \: dx$ per unit length upon use of (\ref{e: L-and-S}) reduces to
	\beq
	P = \Tr \int \ov \la e^{Kx} \dot R \left( [K, R] + mK \right) e^{-Kx} dx = \frac{l}{\la} \Tr \left(S - \ov \la K \right)L = \frac{l k^2}{\la} \left( h - \frac{m}{\la} \right).
	\eeq 
As shown in \S\ref{s:Hamiltonian-mechanical}, the field energy per unit length reduces to the RR model Hamiltonian (\ref{e: H-mechanical}).



Infinitesimal internal rotations $\phi \to \phi + \tht [n, \phi]$ (for $n \in \mathfrak{su}(2)$ and small angle $\tht$) are symmetries of (\ref{e:Lagrangian-scalar-field}) leading to the Noether density and current:
	\beq
	j_t = \Tr \left( \frac{n}{\la} [ \phi, \dot \phi ] -  \frac{n}{3} [\phi, [\phi, \phi']] \right) \quad \text{and} \quad 
	j_x = \Tr \left( -\frac{n}{\la}[ \phi , \phi'] +  \frac{n}{3} [\phi, [\phi, \dot \phi]] \right)
	\label{e:Q-internal-rotation}
	\eeq
and the conservation law $\Tr \left( n \left[ \phi, \frac{\ddot\phi - \phi''}{\la} - [\dot \phi, \phi']\right]\right) = 0$. However, the charges $ Q^{\rm rot}_n = \int j_t \: dx$ do not reduce to conserved quantities of the RR model. This is because the space of mechanical states is {\it not} invariant under the above rotations as $K = i k \sig_3/2$ picks out the third direction.

\subsection{Static and Circular submanifolds}
\label{s:Static-and-trigonometric-solutions}
In general, solutions of the EOM of the RR model (\ref{e: EOM-LS}) are expressible in terms of elliptic functions \cite{R-R}. Here, we discuss the `static' and `circular' (or `trigonometric') submanifolds of the phase space where solutions to (\ref{e: EOM-LS}) reduce to either constant or circular functions of time. Interestingly, these are precisely the places where the conserved quantities fail to be  independent as will be shown in \S \ref{s:independence}.

\subsubsection*{Static submanifolds}

By a static solution on the $L$-$S$ phase space we mean that the six variables $L_a$ and $S_b$ are time-independent. We infer from (\ref{e:EOM-LS-explicit}) that static solutions occur precisely when $S_1 = S_2 = 0$ and $S_3 L_2 = S_3 L_1 = 0$. These conditions lead to two families of static solutions $\Sigma_3$ and $\Sigma_2$. The former is a 3-parameter family defined by $S_{1,2,3}= 0$ with the $L_a$ being arbitrary constants. The latter is a 2-parameter family where $L_3$ and $S_3$ are arbitrary constants while $L_{1,2} = S_{1,2} = 0$. We will refer to $\Sigma_{2,3}$ as `static' submanifolds of $M^6_{S \text{-} L}$. Their intersection is the $L_3$ axis. Note however, that the `extra coordinate' $R_3(t)$ corresponding to such solutions evolves linearly in time, $R_3(t) = R_3(0) + (S_3 + k/\la)t$.

The conserved quantities satisfy interesting relations on $\Sigma_2$ and $\Sigma_3$. On $\Sigma_2$ we must have $h = \mp \sign{k} \: m s$ and ${\mathfrak{c}} = m^2/ 2 \pm \sign{k} \: s/ \la$ with $s \geq 0$ where the signs correspond to the two possibilities $S_3 = \pm s |k|$. Similarly, on $\Sigma_3$ we must have $s = h = 0$ with $2 {\mathfrak{c}} - m^2 \geq 0$. While $\Sigma_3$ may be regarded as the pre-image (under the map introduced in \S \ref{s:conserved-quantities}) of the submanifold $s = 0$ of the space of conserved quantities $\cal Q$, $\Sigma_2$ is {\it not} the inverse image of any submanifold of $\cal Q$. In fact, the pre-image of the submanifold of $\cal Q$ defined by the relations that hold on $\Sigma_2$ also includes many interesting non-static solutions that we shall discuss elsewhere.
\subsubsection*{Circular or Trigonometric submanifold}

As mentioned in \S \ref{s:conserved-quantities} the EOM may be solved in terms of elliptic functions \cite{R-R}. In particular, since from (\ref{e:EOM-u}) $\dot u^2 = 2 \la k^2 \chi(u)$, $u$ oscillates between a pair of adjacent zeros of the cubic $\chi$, between which $\chi > 0$. When the two zeros coalesce $u = S_3/k$ becomes constant in time. From (\ref{e:EOM-LS-explicit}) this implies $S_1 L_2 = S_2 L_1$, which in turn implies that $\tan \tht = \tan \phi$ or $\tht - \phi = n \pi$ for an integer $n$. Moreover, $\rho, r$ and $\dot{\tht} = \dot{\phi}$ become constants as from (\ref{e:theta-phi-dynamics}), they are functions of $u$. Thus the EOM for $S_1 = k \rho \cos \phi$ and $S_2 = k \rho \sin \phi$ simplify to $\dot{S_1} = -\dot{\phi} S_2$ and $\dot{S_2} = \dot{\phi} S_1$ with solutions given by circular functions of time. The same holds for $L_1 = k r \cos \tht$ and $L_2 = k r \sin \tht$ as $\dot{L_1} = k S_2$ and $\dot{L_2} = -k S_1$ (\ref{e:EOM-LS-explicit}). Thus, we are led to introduce the circular submanifold of the phase space as the set on which solutions degenerate from elliptic to circular functions. In what follows, we will express it as an algebraic subvariety of the phase space. Note first, using (\ref{e:L-S-polar}), that on the circular submanifold
	\beq
	\dot{\tht} = \dot{\phi} = (-1)^{n+1}\frac{k \rho}{r} = -\frac{k S_1}{L_1} = -\frac{k S_2}{L_2}.
	\eeq
Thus EOM on the circular submanifold take the form
	\beq
	\dot L_3 = \dot S_3 = 0, \quad
	\dot L_1 = k S_2, \quad
	\dot L_2 = -k S_1, \quad \dot S_1 = \frac{k S_2^2}{L_2}
	\quad  \text{and} \quad
	\dot S_2 = -\frac{k S_1^2}{L_1}.
	\label{e:EOM-trigonometric-submanifold}
	\eeq
The non-singular nature of the Hamiltonian vector field $V_E$ ensures that the above quotients make sense. Interestingly, the EOM (\ref{e:EOM-LS-explicit}) reduce to (\ref{e:EOM-trigonometric-submanifold}) when $S$ and $L$ satisfy the following three relations
	\beq
	\Xi_1: \; (S \times L)_3 = 0, \quad
	\Xi_2: -\la L_1(S \times L)_2 = k S_1^2 
	\quad \text{and} \quad
	\Xi_3: \: \la L_2 (S \times L)_1 = k S_2^2.
	\label{e:trigonometric-submanifold-conditions}
	\eeq
Here $(S \times L)_3 = S_1 L_2 - S_2 L_1$ etc. The conditions (\ref{e:trigonometric-submanifold-conditions}) define a singular subset $\bar {\cal C}$ of the phase space. $\bar {\cal C}$ may be regarded as a disjoint union of the static submanifolds $\Sigma_2$ and $\Sigma_3$ as well as the three submanifolds ${\cal C}$, ${\cal C}_1$ and ${\cal C}_2$ of dimensions four, three and three, defined by:
	\beqs
	{\cal C}: && S_1 \neq 0, \quad S_2 \neq 0, \quad \Xi_1 \;\; \text{and either} \;\; \Xi_2 \;\; \text{or} \;\; \Xi_3 , \cr
	{\cal C}_1: && S_1 =0,\quad S_2 \neq 0, \quad  L_1 = 0  \quad \text{and} \quad \Xi_3 \cr
	\text{and} \quad {\cal C}_2: && S_1 \neq 0,\quad S_2 = 0, \quad  L_2 = 0 \quad \text{and}\quad \Xi_2. \qquad
	\label{e:singular-submanifolds-for-4wedge-vanish}
	\eeqs
${\cal C}_1$, ${\cal C}_2$, $\Sigma_2$ and $\Sigma_3$ lie along boundaries of ${\cal C}$. The dynamics on ${\cal C}$ (where $L_{1,2}$ and $S_{1,2}$ are necessarily non-zero) is particularly simple. We call ${\cal C}$ the circular submanifold, it is an invariant submanifold on which $S$ and $L$ are circular functions of time. Indeed, to solve (\ref{e:EOM-trigonometric-submanifold}) note that the last pair of equations may be replaced with $\dot L_1/L_1 = \dot S_1/S_1$ and $\dot L_2/L_2 = \dot S_2/S_2$ which along with $S_1 L_2 = S_2 L_1$ implies that $S_{1,2} = \al L_{1,2}$ for a constant $\al > 0$. Thus we must have $\dot S_1 = k \al S_2$ and $\dot S_2 = - k \al S_1$ with the solutions
	\beq
	S_1/k = A \sin(k \al t) + B \cos(k \al t) \quad \text{and} 
	\quad S_2/k = - B \sin( k \al t) + A\cos(k \al t).
	\eeq
$A$ and $B$ are dimensionless constants of integration. As a consequence of $\Xi_2$ or $\Xi_3$ (\ref{e:trigonometric-submanifold-conditions}), the constant values of $L_3 = -k m$ and $S_3 = u k$ must satisfy the relation $u = -\alpha (\alpha + \la m)/\la$. The other conserved quantities are given by 
	\beqs
	{\mathfrak{c}} &=& \frac{1}{2} \left(m^2 + 
	\frac{A^2 + B^2}{\al^2} - \frac{2 \al (\al + \la m)}{\la^2} \right),
	\quad
	 h = \frac{A^2+B^2}{\al} + \frac{\al m (\al +\la m)}{\la} \quad \text{and} \cr
	 s^2 &=& A^2+B^2+\frac{\al^2 (\al + \la m)^2}{\la^2}.
	 \eeqs
Though we do not discuss it here, it is possible to show that these trigonometric solutions occur precisely when the common level set of the four conserved quantities is a circle as opposed to a 2-torus. Unlike $\Sigma_2$ and $\Sigma_3$, the boundaries ${\cal C}_1$ and ${\cal C}_2$ are {\it not} invariant under the dynamics. The above trajectories on ${\cal C}$ can reach points of ${\cal C}_1$ or ${\cal C}_2$, say when $S_1$ or $S_2$ vanishes. On the other hand, in the limit $A = B = 0$ and $\al \neq 0$, the above trigonometric solutions reduce to the $\Sigma_2$ family of static solutions. What is more, $\Sigma_2$ lies along the common boundary of ${\cal C}_1$ and ${\cal C}_2$. Finally, when $A$, $B$ and $\al$ are all zero, $S_1, S_2$ and $S_3$ must each vanish while $L_1, L_2$ and $L_3$ are arbitrary constants. In this case, the trigonometric solutions reduce to the $\Sigma_3$ family of static solutions.

\subsection{Independence of conserved quantities and singular submanifolds} 
\label{s:independence}

We wish to understand the extent to which the above four conserved quantities are independent. We say that a pair of conserved quantities, say $f$ and $g$, are independent if $df$ and $dg$ are linearly independent or equivalently  if $df \wedge dg$ is not identically zero. Similarly, three conserved quantities are independent if $df \wedge dg \wedge dh \not \equiv 0$ and so on. In the present case, we find that the pairwise, triple and quadruple wedge products of $d{\mathfrak{c}}, dh, dm$ and $ds^2$ do not vanish identically on the whole $L$-$S$ phase space. Thus the four conserved quantities are generically independent. However, there are some `singular' submanifolds of the phase space where these wedge products vanish and relations among the conserved quantities emerge. This happens precisely on the static submanifolds $\Sigma_{2,3}$ and $\bar {\cal C}$ which includes the circular submanifold and its boundaries discussed in \S\ref{s:Static-and-trigonometric-solutions}.

A related question is the independence of the canonical vector fields obtained through contraction of the 1-forms with the (say, nilpotent) Poisson tensor $\scripty{r}_0$. The Casimir vector fields $V_{\mathfrak{c}}$ and $V_m$ are identically zero as $d{\mathfrak{c}}$ and $dm$ lie in the kernel of $\scripty{r}_0$. Passing to the symplectic leaves $M^4_{{\mathfrak{c}} m}$, we find that the vector fields corresponding to the non-Casimir conserved quantities $V_E$ and $V_h$ are generically linearly independent.  Remarkably, this independence fails precisely where $M^4_{{\mathfrak{c}} m}$ intersects $\bar {\cal C}$.

\subsubsection*{Conditions for pairwise independence of conserved quantities}
 
The 1-forms corresponding to our four conserved quantities are
	\beq
	k^2 ds^2 = 2 S_a dS_a, \quad k^2 d{\mathfrak{c}} = L_a dL_a + \frac{k}{\la} dS_3, \quad
	-k \: dm = dL_3 \quad \text{and} \quad  k^2 dh = S_a dL_a + L_a dS_a.
	\eeq
None of the six pairwise wedge products is identically zero:
	\beqs
	\frac{k^4}{2} ds^2 \wedge dh &=& S_a S_b dS_a \wedge dL_b + \half (S_a L_b - S_b L_a) dS_a \wedge dS_b, 
	\quad
	\frac{k^3}{2} dm \wedge ds^2  = S_a dS_a \wedge dL_3 \cr
	k^3 dm \wedge dh &=& S_a dL_a \wedge dL_3 + L_a dS_a \wedge dL_3, 
	\quad  
	k^3 d{\mathfrak{c}} \wedge dm = L_a dL_3 \wedge dL_a + \frac{k}{\la} dL_3 \wedge dS_3 \cr 
	\frac{k^4}{2} ds^2 \wedge d{\mathfrak{c}} &=&  S_a L_b dS_a \wedge dL_b + \frac{kS_a}{\la} dS_a \wedge dS_3 \cr
	k^4 dh \wedge d{\mathfrak{c}} &=& \half(S_a L_b - S_b L_a) dL_a \wedge dL_b - \sum _{b \neq 3} L_a L_b dL_a \wedge dS_b + \frac{k L_a}{\la} dS_a \wedge dS_3 \cr
	&& + \left( \frac{k S_a}{\la} - L_a L_3 \right) dL_a \wedge dS_3.  
	\eeqs
Though no pair of conserved quantities is dependent on $M^6_{S \text{-} L}$, there are some relations between them on certain submanifolds. For instance, $ds^2 \wedge dh = ds^2 \wedge dm = 0$ on the $3$D submanifold $\Sigma_3$ (where $s = 0$) while $dh \wedge dm = 0$ on the curve defined by $S_{1,2} = L_{1,2,3} = 0$ where $h = m = 0$. Similarly, $ds^2 \wedge d{\mathfrak{c}} = 0$ on both these submanifolds where $s = 0$ and $\la^2 {\mathfrak{c}}^2 = k^2 s^2$ respectively. Moreover, $dh \wedge d{\mathfrak{c}} = 0$ on the curve defined by $S_{1,2} = L_{1,2} = L_3^2 - k S_3 / \la = 0$ where $k^2 h^2 = \la^2 {\mathfrak{c}}^3$. However, the dynamics on each of these submanifolds is trivial as each of their points represents a static solution. On the other hand, the Casimirs $m$ and ${\mathfrak{c}}$ are independent on all of $M^6_{S \text{-} L}$ provided $1/\la k^2 \ne 0$.

\subsubsection*{Conditions for relations among triples of conserved quantities:}
 
The four possible wedge products of three conserved quantities are given below.
	\beqs
	\frac{k^5}{2} dh \wedge ds^2 \wedge dm &=& S_a S_b dS_a \wedge dL_b \wedge dL_3 + \half (S_a L_b - S_b L_a) dS_a \wedge dS_b \wedge dL_3 \cr 
	\frac{k^6}{2} ds^2 \wedge dh \wedge d{\mathfrak{c}} &=& \half S_a (S_b L_c - S_c L_b) dS_a \wedge dL_b \wedge dL_c + (S_1 L_2 - S_2 L_1) \frac{k}{\la} dS_1 \wedge dS_2 \wedge dS_3 \cr
	&& + \left[(S_a L_3 - S_3 L_a)L_c - \frac{S_a S_c k}{\la} \right] dS_a \wedge dS_3 \wedge dL_3  \cr
	&& + \sum_{a, b \neq 3} \half( S_a L_b - S_b L_a) L_c dS_a \wedge dS_b \wedge dL_c \cr 
	\frac{k^5}{2} dm \wedge ds^2 \wedge d{\mathfrak{c}} &=& S_a L_b dS_a \wedge dL_3 \wedge dL_b + \frac{k S_a}{\la} dS_a \wedge dL_3 \wedge dS_3 \cr
	k^5 dm \wedge dh \wedge d{\mathfrak{c}} &=& (S_2 L_1 - S_1 L_2) dL_1 \wedge dL_2 \wedge dL_3 + \left( \frac{k S_a}{\la} - L_a L_3 \right) dL_a \wedge dL_3 \wedge dS_3 \cr
	&& - \sum_{b \neq 3} L_a L_b dL_a \wedge dL_3 \wedge dS_b + \frac{k L_a}{\la} dS_a \wedge dL_3 \wedge dS_3.
	\eeqs
It is clear that none of the triple wedge products is identically zero, so that there is no relation among any three of the conserved quantities on all of $M^6_{S \text{-} L}$. However, as before, there are relations on certain submanifolds. For instance, $ds^2 \wedge dm \wedge d{\mathfrak{c}} = ds^2 \wedge dh \wedge d{\mathfrak{c}} = ds^2 \wedge dh \wedge dm = 0$ on both the static submanifolds $\Sigma_3$ and $\Sigma_2$ of \S \ref{s:Static-and-trigonometric-solutions}. On $\Sig_2$ we have the three relations $s^2 = (\la^2/4) (2{\mathfrak{c}} - m^2)^2$, $\la^2 (2{\mathfrak{c}} s^2 - h^2)^2 = 4s^6$ and $h^2 = m^2 s^2$. On the other hand, $dh \wedge dm \wedge d{\mathfrak{c}} = 0$ only on the static submanifold $\Sigma_2$ on which the relation $4h^2 = \la^2 m^2 (2{\mathfrak{c}} - m^2)^2$ holds. 

\subsubsection*{Vanishing of four-fold wedge product and the circular submanifold}

Finally, the wedge product of all four conserved quantities is
	\beqs
	\frac{k^7}{2}dh \wedge ds^2 \wedge dm \wedge d{\mathfrak{c}} &=& (S_1 L_2 - S_2 L_1)\bigg[S_b dL_1 \wedge dL_2 \wedge dL_3 \wedge dS_b \cr 
	&& - \frac{k}{\la} dS_1 \wedge dS_2 \wedge dS_3 \wedge dL_3 - L_b dS_1 \wedge dS_2 \wedge dL_b \wedge dL_3 \bigg] \cr 
	&& + \left[\frac{S_a S_b k}{\la} + (L_a S_3 - S_a L_3) L_b\right] dS_a \wedge dS_3 \wedge dL_b \wedge dL_3.
	\label{e:four-fold-wedge-product}
	\eeqs
This wedge product is {\it not} identically zero on the $L$-$S$ phase space so that the four conserved quantities are independent in general. It does vanish, however, on the union of the two static submanifolds $\Sigma_2$ and $\Sigma_3$. This is a consequence, say, of $ds^2 \wedge dm \wedge d{\mathfrak{c}}$ vanishing on both these submanifolds. Alternatively, if $S_1 = S_2 = 0$, then requiring $ dh \wedge ds^2 \wedge dm \wedge d{\mathfrak{c}} = 0$ implies either $S_3 = 0$ or $L_1 = L_2 = 0$. Interestingly, the four-fold wedge product also vanishes elsewhere. In fact, the necessary and sufficient conditions for it to vanish are $\Xi_1, \Xi_2$ and $\Xi_3$ introduced in (\ref{e:trigonometric-submanifold-conditions}) which define the submanifold $\bar {\cal C}$ of the phase space that includes the circular submanifold ${\cal C}$ and its boundaries ${\cal C}_{1,2}$ and $\Sig_{2,3}$. 
 
Consequent to the vanishing of the four-fold wedge product $dh \wedge ds^2 \wedge dm \wedge d{\mathfrak{c}}$, the conserved quantities must satisfy a new relation on ${\cal C}$ which may be shown to be the vanishing of the discriminant $\D ({\mathfrak{c}}, m , s^2, h)$ of the cubic polynomial 
	\beq
	\chi(u) = u^3 - \la {\mathfrak{c}} u^2 - \left( s^2 + \la h m \right) u + \frac{\la}{2}\left( 2 {\mathfrak{c}}s^2 - h^2 - m^2 s^2 \right). 
	\label{e:cubic-equation-S3}
	\eeq
The properties of $\chi$ help to characterize the common level sets of the four conserved quantities. In fact, $\chi$ has a double zero when the common level set of the four conserved quantities is a circle (as opposed to a 2-torus) so that it is possible to view ${\cal C}$ as a union of circular level sets. Note that $\D$ in fact vanishes on a submanifold of phase space that properly contains $\bar {\cal C}$. However, though the conserved quantities satisfy a relation on this larger submanifold, their wedge product only vanishes on $\bar {\cal C}$. The nature of the common level sets of conserved quantities will be examined elsewhere.

\subsubsection*{Independence of Hamiltonian and helicity on symplectic leaves $M^4_{{\mathfrak{c}}m}$}

So far, we examined the independence of conserved quantities on $M^6_{S \text{-} L}$ which, however, is a degenerate Poisson manifold. By assigning arbitrary real values to the Casimirs $\mathfrak{c}$ and $m$ (of $\{ \cdot , \cdot \}_\nu$) we go to its symplectic leaves $M^4_{{\mathfrak{c}} m}$. $L_{1,2}$ and $S_{1,2}$ furnish coordinates on $M^4_{{\mathfrak{c}} m}$ with
	\beq
	S_3 (L_1, L_2) =  \frac{\la k}{2}\left( (2 {\mathfrak{c}} - m^2) - \frac{1}{k^2} (L_1^2 + L_2^2) \right)  \quad \text{and} \quad L_3 = -m k.
	\label{e:S3-and-L3-on-M4-cm}
	\eeq
The Hamiltonian $H = E k^2$ (or $k^2 s^2 = 2 (H - {\mathfrak{c}} k^2 - k^2/ 2 \la^2) $) and helicity $h$ are conserved quantities for the dynamics on $M^4_{{\mathfrak{c}} m}$. Here we show that the corresponding vector fields $V_E$ and $V_h$ are generically independent on each of the symplectic leaves and also identify where the independence fails. On $M^4_{{\mathfrak{c}} m}$, the Poisson tensor $\scripty{r}_0$ is nondegenerate so that $V_E$ and $V_h$ are linearly independent iff $dE \wedge dh \neq 0$. We find 
	\beqs
	k^5 dE \wedge dh &=& (S_1 L_2 - S_2 L_1)\left( k dS_1 \wedge dS_2 + \la S_3 dL_1 \wedge dL_2 \right) \cr
	&& + \sum_{a,b =1,2}\left(\la (S_b L_3 - S_3 L_b)L_a - k S_a S_b \right) dL_a \wedge dS_b
	\eeqs 
Here $S_3$ and $L_3$ are as in (\ref{e:S3-and-L3-on-M4-cm}). Interestingly, the conditions for $dE \wedge dh$ to vanish are the same as the restriction to $M^4_{{\mathfrak{c}}m}$ of the conditions for the vanishing of the four-fold wedge product $dh \wedge ds^2 \wedge dm \wedge d{\mathfrak{c}}$ (\ref{e:four-fold-wedge-product}).  It is possible to check that this wedge product vanishes on $M^4_{{\mathfrak{c}} m}$ precisely when $S_{1,2}$ and $L_{1,2}$ satisfy the relations $\Xi_1, \Xi_2$ and $\Xi_3$ of (\ref{e:trigonometric-submanifold-conditions}), where $S_3$ (\ref{e:S3-and-L3-on-M4-cm}) and $L_3 = -m k$ are expressed in terms of the coordinates on $M^4_{{\mathfrak{c}}m}$. Recall from \S \ref{s:Static-and-trigonometric-solutions} that (\ref{e:trigonometric-submanifold-conditions}) is satisfied on the singular set $\bar {\cal C} \subset M^6_{S \text{-} L}$ consisting of the union of the circular submanifold ${\cal C}$ and its boundaries ${\cal C}_{1,2}$ and $\Sigma_{2,3}$. Thus, on $M^4_{{\mathfrak{c}} m}$ $V_E$ and $V_h$ are linearly independent away from the set (of measure zero) given by the intersection of $\bar {\cal C}$ with $M^4_{{\mathfrak{c}} m}$. For example, the intersections of ${\cal C}$ with $M^4_{{\mathfrak{c}} m}$ are in general $2$D manifolds defined by four conditions among $S$ and $L$: $\Xi_1$ and $\Xi_2$ (with $S_{1,2} \neq 0$) as well as the condition (\ref{e:S3-and-L3-on-M4-cm}) on $S_3$ and finally $L_3 = -m k$. {\it This independence along with the involutive property of $E$ and $h$ allows us to conclude that the system is Liouville integrable on each of the symplectic leaves.}

We note in passing that the $E$ and $h$ when regarded as functions on $M^6_{S \text{-} L}$ (rather than $M^4_{{\mathfrak{c}} m}$) are independent everywhere except on a curve that lies on the static submanifold $\Sigma_2$. In fact, we find that $dE \wedge dh$ vanishes iff $S_{1,2} = L_{1,2} = 0$ and $S_3^2 + kS_3/ \la = L_3^2$.

\section{Similarities and differences with the Neumann model}
\label{s:RR-model-and-Neumann-Model}

The EOM (\ref{e: EOM-LS}) and Lax pair (\ref{e:Lax-pair}) of the RR model have a formal structural similarity with those of the Neumann model. The latter describes the motion of a particle on $S^{N-1}$ subject to harmonic forces with frequencies $a_1, \cdots, a_N$ \cite{B-B-T}. In other words, a particle moves on  $S^{N-1} \subset \mathbb{R}^N$ and is connected by $N$ springs, the other ends of which are free to move on the $N$ coordinate hyperplanes. The EOM of the Neumann model follow from a symplectic reduction of dynamics on a $2N$ dimensional phase space with coordinates $x_1, \cdots, x_N$ and $y_1, \cdots, y_N$. The canonical PBs $\{ x_k, y_l \} = \del_{kl}$ and Hamiltonian
	\beq
	H = \frac{1}{4} \sum_{k \neq l} J_{kl}^2 + \half \sum_{k} a_k x_k^2 
	\label{e:Hamiltonian-Neumann}
	\eeq
lead to Hamilton's equations 
	\beq
	\dot x_k = - J_{kl} x_l \quad \text{and} \quad \dot y_k = - J_{kl} y_l - a_k x_k \quad (\text{no sum over $k$}).
	\eeq
Here, $J_{kl} = x_k y_l - x_l y_k$ is the angular momentum. Introducing the column vectors $X_k = x_k$ and $Y_k = y_k$ and the frequency matrix $\Om = \text{diag} (a_1, \cdots, a_N)$, Hamilton's equations become
	\beq
	\dot X = - J X \quad \text{and} \quad \dot Y = - J Y - \Om X.
	\eeq
It is easily seen that $X^t X$ is a constant of motion. Moreover, the Hamiltonian and PBs are invariant under the `gauge' transformation $(X, Y) \to (X , Y + \eps X)$ for $\eps \in \mathbb{R}$. Imposing the gauge condition $X^t (Y + \eps(t) X) = 0$ along with $X^t X = 1$ allows us to reduce the dynamics to a phase space of dimension $2(N-1)$. If we define the rank 1 projection $P = X X^t$ then $J = X Y^t - Y X^t$ and $P$ are seen to be gauge-invariant and satisfy the evolution equations
	\beq
	\dot J = [\Om, P] \quad \text{and} \quad \dot P = [P, J].
	\label{e:EOM-Neumann-angmom-proj}
	\eeq
The Hamiltonian (\ref{e:Hamiltonian-Neumann}) in terms of $J, P$ and $\Om$ becomes
	\beq
	H_{\rm Neu} =  \tr\left(-\frac{1}{4} J^2 + \half \Om P \right).
	\label{e:Hamiltonian-Neumann-JP} 
	\eeq
The PBs following from the canonical $x$-$y$ PBs 
	\beqs
	\{ J_{kl} , J_{pq} \} &=& \del_{kq} J_{pl} - \del_{pl} J_{kq}  + \del_{ql} J_{kp}  - \del_{kp} J_{ql}, \cr
	\{ P_{kl} , J_{pq} \} &=& \del_{kq} P_{pl} - \del_{pl} P_{kq}  + \del_{ql} P_{kp}  - \del_{kp} P_{ql} \;\; \text{and} \;\; \{ P_{kl} , P_{pq} \} = 0
	\label{e:PB-JP-Neumann}
	\eeqs
and the Hamiltonian (\ref{e:Hamiltonian-Neumann-JP}) imply the EOM (\ref{e:EOM-Neumann-angmom-proj}). This Euclidean Poisson algebra is a semi-direct product of the abelian ideal spanned by the $P$'s and the simple Lie algebra of the $J$'s.

Notice the structural similarity between the equations of the RR model (\ref{e: EOM-LS}) and those of the Neumann model (\ref{e:EOM-Neumann-angmom-proj}). Indeed, under the mapping $(L, S, K, \la) \mapsto (J, P, \Om, 1)$, the EOM (\ref{e: EOM-LS}) go over to (\ref{e:EOM-Neumann-angmom-proj}). The Lax pair for the Neumann model \cite{B-B-T}
	\beq
	L(\zeta) = -\Om + \ov \zeta J + \ov{\zeta^2} P \quad \text{and} \quad M(\zeta) = \ov \zeta P \quad \text{with} \quad \dot L = [M, L]
	\eeq
and that of the RR model $A_{\varepsilon}(\zeta) = -K + L/\zeta + S/ (\la \zeta^2)$ and $B(\zeta) = S/ \zeta$ (\ref{e:Lax-pair}) are similarly related for $\la = 1$. Despite these similarities, there are significant differences. 

(a) While $L$ and $S$ are Lie algebra-valued traceless anti-hermitian matrices, $J$ and $P$ are a real anti-symmetric and a real symmetric rank-one projection matrix. Furthermore, while $K$ is a constant traceless anti-hermitian matrix ($(ik/2) \sig_3$ for $\mathfrak{su}(2)$), the frequency matrix $\Om$ is diagonal with positive entries. 

(b) The Hamiltonian (\ref{e:Hamiltonian-Neumann-JP}) of the Neumann model also differs from that of our model (\ref{e: H-mechanical}) as it does not contain a quadratic term in $P$. However, the addition of $(1/4)\tr P^2$ to (\ref{e:Hamiltonian-Neumann-JP}) would not alter the EOM (\ref{e:EOM-Neumann-angmom-proj}) as $\tr P^2$ is a Casimir of the algebra (\ref{e:PB-JP-Neumann}).

(c) The PBs (\ref{e:PB-JP-Neumann}) of the Neumann model bear some resemblance to the Euclidean PBs (\ref{e:PB-semi-diect-tilde-LS}) of the RR model expressed in terms of the real anti-symmetric matrices $\tl S$ and $\tl L$ of \S \ref{s:Hamiltonian-mechanical}. Under the map $(\tl L, \tl S, \la) \mapsto (J, P, 1)$, the PBs (\ref{e:PB-semi-diect-tilde-LS}) go over to (\ref{e:PB-JP-Neumann}) up to an overall factor of $-1/2$. On the other hand, if we began with the $\{ \tl L_{kl}, \tl S_{pq} \}_{\varepsilon}$ PB implied by (\ref{e:PB-semi-diect-tilde-LS}) and then applied the map, the resulting $\{ J, P \}$ PB would be off by a couple of signs. These sign changes are necessary to ensure that the $J$-$P$ PBs respect the symmetry of $P$ as opposed to the anti-symmetry of $\tl S$. This also reflects the fact that the symmetry $\{ \tl S_{kl}, \tl L_{pq} \} =\{ \tl L_{kl}, \tl S_{pq} \}$ is not present in the Neumann model: $\{ J_{kl}, P_{pq} \} \neq \{ P_{kl}, J_{pq} \}$.

(d) Though both models possess non-dynamical $r$-matrices, they are somewhat different as are the forms of the fundamental PBs among Lax matrices. Recall that the FPBs and $r$-matrix (\ref{e:r-matrix-semi-direct}) of the RR model, say, for the Euclidean PBs are (here, $k,l,p,q = 1,2$):
	\beq
	\left\{ A_{\varepsilon}(\zeta) \stackrel{\otimes}{,} A_{\varepsilon}(\zeta') \right\}_{\varepsilon} = \left[ r_{\varepsilon}(\zeta, \zeta') , A_{\varepsilon}(\zeta)\otimes I + I \otimes A_{\varepsilon}(\zeta') \right] \;\; \text{and} \;\; r_{\varepsilon}(\zeta, \zeta')_{k l p q} = -\frac{\la \: \del_{k q} \del_{lp}}{2 (\zeta - \zeta')}.
	\eeq
This $r$-matrix has a single simple pole at $\zeta = \zeta'$. On the other hand, the FPBs of the Neumann model may be expressed as a sum of {\it two} commutators
	\beq
	\{ L(\zeta) \stackrel{\otimes}{,} L(\zeta') \} = [ r_{12}(\zeta, \zeta'), L(\zeta) \otimes I ] - [ r_{21}(\zeta', \zeta) , I \otimes L(\zeta') ].
	\label{e:FPB-Neumann}
	\eeq
The corresponding $r$-matrices have simple poles at $\zeta = \pm \zeta'$ (here, $k,l,p,q = 1, \cdots, N$):
	\beq
	r_{12}(\zeta, \zeta')_{klpq} = -\frac{\del_{kq} \del_{lp}}{\zeta - \zeta'}  - \frac{\del_{kl} \del_{pq}}{\zeta + \zeta'}  
	\quad \text{and} \quad
	r_{21}(\zeta', \zeta)_{klpq} = -\frac{\del_{kq} \del_{lp}}{\zeta' - \zeta}  - \frac{\del_{kl} \del_{pq}}{\zeta' + \zeta} \neq -r_{12}(\zeta, \zeta')_{klpq}.
	\eeq	
Note that the anti-symmetry of (\ref{e:FPB-Neumann}) is guaranteed by the relation $r_{12}(\zeta, \zeta')_{klpq} = r_{21}(\zeta,\zeta')_{lkqp}$.
	

\vspace{.25cm}

{\fl \bf New Hamiltonian formulation for the Neumann model:} An interesting consequence of our analogy is a new Hamiltonian formulation for the Neumann model inspired by the nilpotent RR model PBs (\ref{e:PB-nilpotent-tilde-LS}). Indeed, suppose we take the Hamiltonian for the Neumann model as 
	\beq
	H = H_{\rm Neu} + \ov{4} \tr P^2 = \tr \left( -\ov{4} J^2 + \half \Omega P + \ov{4} P^2 \right)
	\eeq
and postulate the step-3 nilpotent PBs, 
	\beqs
	\{ P_{kl}, J_{pq} \}_{\nu} &=& - \del_{kq} \Omega_{pl} + \del_{pl} \Omega_{kq} - \del_{ql} \Omega_{kp} + \del_{kp} \Omega_{ql}, \cr
	\{ P_{kl} , P_{pq} \}_{\nu} &=& \del_{kq} J_{pl} - \del_{pl} J_{kq} - \del_{ql} J_{kp} + \del_{kp} J_{ql} \quad \text{and} \quad \{ J_{kl} , J_{pq} \}_{\nu} = 0,
	\label{e:nilpotent-PB-Neumann}
	\eeqs
then Hamilton's equations reduce to the EOM (\ref{e:EOM-Neumann-angmom-proj}). These PBs differ from those obtained from (\ref{e:PB-nilpotent-tilde-LS}) via the map $(\tl L, \tl S, \tl K, \la) \mapsto  (J, P, \Omega, 1)$ by a factor of $1/2$ and a couple of signs in the $\{P, P \}_{\nu}$ PB. As before, these sign changes are necessary since $P$ is symmetric while $\tl S$ is anti-symmetric. It is straightforward to verify that the Jacobi identity is satisfied: the only non-trivial case being $\{ \{ P, P \}, P \} + \rm{cyclic} = 0$ where cancellations occur among the cyclically permuted terms. In all other cases the individual PBs such as $\{ \{ P, J \}, J \}$ are identically zero. Though inspired by the $\mathfrak{su}(2)$ case of the RR model, the PBs (\ref{e:nilpotent-PB-Neumann}) are applicable to the Neumann model for all values of $N$.

\section{Discussion}
\label{s:Discussion}

In this paper, we studied the classical Rajeev-Ranken model which is a mechanical reduction of a nilpotent scalar field theory dual to the 1+1-dimensional SU(2) principal chiral model. We find a Lagrangian as well as a pair of distinct Hamiltonian-Poisson bracket formulations for this model. The corresponding nilpotent and Euclidean Poisson brackets are shown to be compatible and to generate a (degenerate) Poisson pencil. Lax pairs and $r$-matrices associated with both Poisson structures are obtained and used to find four generically independent conserved quantities which are in involution with respect to either Poisson structure on the six-dimensional phase space, thus indicating the Liouville integrability of the model. The symmetries and canonical transformations generated by these conserved quantities are identified and three of their combinations are related to Noether charges of the nilpotent scalar field theory. Two of these conserved quantities (${\mathfrak{c}}$ and $m$ or $s$ and $h$) are shown to lie in the centers of the corresponding Poisson algebras. Thus, by assigning numerical values to the Casimirs we may go from the 6D phase space of the model to its 4D symplectic leaves $M^4_{{\mathfrak{c}} m}$ or $M^4_{s h}$ on which we have two generically independent conserved quantities in involution, thereby rendering the system Liouville integrable. Though all four conserved quantities are shown to be generically independent, there are singular submanifolds of the phase space where this independence fails. In fact, we find the submanifolds where pairs, triples or all four conserved quantities are dependent and identify the relations among conserved quantities on them. Remarkably, these submanifolds are shown to coincide with the `static' and `circular/trigonometric' submanifolds of the phase space and to certain non-generic common level sets of conserved quantities. 

As an unexpected payoff from our study of the algebraic structures of the RR model, we find a new Hamiltonian formulation for the Neumann model. Though we find that the equations of motion, Hamiltonians and Lax pairs of the models are formally related, their phase spaces, Poisson structures and $r$-matrices differ in interesting ways. 

Though we have argued that the RR model is Liouville integrable, it remains to explicitly identify action-angle variables on the phase space. It is also of interest to find all common level sets of conserved quantities and describe the foliation of the phase space by invariant tori of various dimensions. The possible extension of the algebraic structures and integrability of this mechanical reduction to its quantum version and its parent nilpotent scalar field theory is of course of much interest. We intend to address these issues in future work.

\vspace{.25cm}

{\fl \bf Acknowledgements:} We would like to thank S G Rajeev for getting us interested in this model and also thank G Date and V V Sreedhar for useful discussions. This work was supported in part by the Infosys Foundation.

\appendix
\section{Calculation of $\Tr A^4(\zeta)$ for the Lax matrix}
\label{a:A^4}

In \S \ref{s:conserved-quantities} we found that the conserved quantities $\Tr A^n(\zeta)$ are in involution and obtained four independent conserved quantities ${\mathfrak{c}}, m, s$ and $h$ by taking $n = 2$. Here, we show that the conserved quantities following from $\Tr A^4(\zeta)$ are functions of the latter. We find that 
\footnotesize
	\beqs
	A^4 &=&  \bigg[ \zeta^8 (K_a K_b K_c K_d) 
				  - \zeta^7 (K_a K_b K_c L_d + L_a K_b K_c K_d + K_a L_b K_c K_d + K_a K_b L_c K_d) \cr
			   &+&  \zeta^6 \left(- \frac{K_a K_b K_c S_d}{\la} + L_a K_b K_c L_d + K_a L_b K_c L_d + K_a K_b L_c L_d \right. \cr  && \left.- \frac{S_a K_b K_c K_d}{\la} + L_a L_b K_c K_d - \frac{K_a S_b K_c K_d}{\la} + L_a K_b L_c K_d + K_a L_b L_c K_d - \frac{K_a K_b S_c K_d}{\la} \right) \cr
			   &+&  \zeta^5 \left(\frac{L_a K_b K_c S_d}{\la} + \frac{K_a L_b K_c S_d}{\la} + \frac{K_a K_b L_c S_d}{\la} \right. \cr && + \left. \frac{S_a K_b K_c L_d}{\la} - L_a L_b K_c L_d + \frac{K_a S_b K_c L_d}{\la} - L_a K_b L_c L_d - K_a L_b L_c L_d + \frac{K_a K_b S_c L_d}{\la} \right.\cr && + \left. \frac{S_a L_b K_c K_d}{\la} + \frac{L_a S_b K_c K_d}{\la} + \frac{S_a K_b L_c K_d}{\la} + \frac{K_a S_b L_c K_d}{\la} + \frac{L_a K_b S_c K_d}{\la} + \frac{K_a L_b S_c K_d}{\la} - L_a L_b L_c K_d\right) \cr
	           &+& \zeta^4 \left(\frac{S_a K_b K_c S_d}{\la^2} - \frac{L_a L_b K_c S_d}{\la} + \frac{K_a S_b K_c S_d}{\la^2} - \frac{L_a K_b L_c S_d}{\la} - \frac{K_a L_b L_c S_d}{\la} + \frac{K_a K_b S_c S_d}{\la^2} \right.\cr && - \left. \frac{S_a L_b K_c L_d}{\la} - \frac{L_a S_b K_c L_d}{\la} - \frac{S_a K_b L_c L_d}{\la} - \frac{K_a S_b L_c L_d}{\la} - \frac{L_a K_b S_c L_d}{\la} - \frac{K_a L_b S_c L_d}{\la} + L_a L_b L_c L_d \right. \cr && \left.\frac{S_a S_b K_c K_d}{\la^2} - \frac{S_a L_b L_c K_d}{\la} - \frac{L_a S_b L_c K_d}{\la} + \frac{S_a K_b S_c K_d}{\la^2} - \frac{L_a L_b S_c K_d}{\la} + \frac{K_a S_b S_c K_d}{\la^2}\right)\cr
	           &+& \zeta^3 \left(- \frac{S_a L_b K_c S_d}{\la^2} - \frac{L_a S_b K_c S_d}{\la^2} - \frac{S_a K_b L_c S_d}{\la^2} - \frac{K_a S_b L_c S_d}{\la^2} - \frac{L_a K_b S_c S_d}{\la^2} - \frac{K_a L_b S_c S_d}{\la^2} + \frac{L_a L_b L_c S_d}{\la} \right.\cr && - \left.\frac{S_a S_b K_c L_d}{\la^2} + \frac{S_a L_b L_c L_d}{\la} + \frac{L_a S_b L_c L_d}{\la} - \frac{S_a K_b S_c L_d}{\la^2} + \frac{L_a L_b S_c L_d}{\la} \right.\cr && - \left. \frac{K_a S_b S_c L_d}{\la^2} - \frac{S_a S_b L_c K_d}{\la^2} - \frac{S_a L_b S_c K_d}{\la^2} - \frac{L_a S_b S_c K_d}{\la^2}\right)\cr
	           &+& \zeta^2 \left(- \frac{S_a S_b K_c S_d}{\la^3} + \frac{S_a L_b L_c S_d}{\la^2} + \frac{L_a S_b L_c S_d}{\la^2} - \frac{S_a K_b S_c S_d}{\la^3} + \frac{L_a L_b S_c S_d}{\la^2} - \frac{K_a S_b S_c S_d}{\la^3} \right.\cr && + \left. \frac{S_a S_b L_c L_d}{\la^2} + \frac{S_a L_b S_c L_d}{\la^2} + \frac{L_a S_b S_c L_d}{\la^2} - \frac{S_a S_b S_c K_d}{\la^3}\right) \cr 
	           &+& \zeta \left(\frac{S_a S_b L_c S_d}{\la^3} + \frac{S_a L_b S_c S_d}{\la^3} + \frac{L_a S_b S_c S_d}{\la^3} + \frac{S_a S_b S_c L_d}{\la^3}\right) + \frac{S_a S_b S_c S_d}{\la^4} \bigg] t_a t_b t_c t_d.
	\eeqs 
\normalsize
Evaluating the trace yields the polynomial (\ref{e:trace-A4}) whose coefficients are functions of the conserved quantities ${\mathfrak{c}}, m , s$ and $h$, thus showing that $\Tr A^4$ does not lead to any new conserved quantity.



\end{document}